\documentclass[floats,floatfix,showpacs,amssymb,prd,superscriptaddress,nofootinbib,twocolumn,aps]{revtex4-1}
\usepackage{graphicx, epsfig, amssymb} 
\usepackage{amsmath, amsfonts}
\usepackage{bm} 

\usepackage[final,breaklinks]{hyperref}
\usepackage[caption=false]{subfig}
\usepackage[usenames]{color}

\def\be{\begin{equation}}
\def\ee{\end{equation}}
\def\beq{\begin{eqnarray}}
\def\eeq{\end{eqnarray}}

\def\p{\partial}

\newcommand{\bea}{\begin{eqnarray}}
\newcommand{\eea}{\end{eqnarray}}
\newcommand{\ben}{\begin{enumerate}}
\newcommand{\een}{\end{enumerate}}
\newcommand{\bi}{\begin{itemize}}
\newcommand{\ei}{\end{itemize}}

\newcommand{\nn}{\nonumber}

\begin{document}

\title{\large Eddington-inspired Born-Infeld gravity. \\ Phenomenology of non-linear gravity-matter coupling 
}

  \author{Paolo Pani}
 \affiliation{CENTRA, Departamento de F\'{\i}sica, 
 Instituto Superior T\'ecnico, Universidade T\'ecnica de Lisboa - UTL,
 Av.~Rovisco Pais 1, 1049 Lisboa, Portugal.}
%
 \author{T\'erence Delsate}
 \affiliation{CENTRA, Departamento de F\'{\i}sica, 
 Instituto Superior T\'ecnico, Universidade T\'ecnica de Lisboa - UTL,
 Av.~Rovisco Pais 1, 1049 Lisboa, Portugal.} 
 \affiliation{Theoretical and Mathematical Physics Dpt.,Universit\'{e} de Mons, UMons, 20, Place du Parc 7000 Mons, Belgium.}
%
 \author{Vitor Cardoso}
 \affiliation{CENTRA, Departamento de F\'{\i}sica, 
 Instituto Superior T\'ecnico, Universidade T\'ecnica de Lisboa - UTL,
 Av.~Rovisco Pais 1, 1049 Lisboa, Portugal.}
 \affiliation{Department of Physics and Astronomy, The University of Mississippi, University, MS 38677, USA.}

\begin{abstract}
Viable corrections to the matter sector of Poisson's equation may result in qualitatively different astrophysical phenomenology, for example the gravitational collapse and the properties of compact objects can change drastically.
We discuss a class of modified non-relativistic theories and focus on a relativistic completion, Eddington-inspired Born-Infeld gravity.
This recently proposed theory is equivalent to General Relativity in vacuum, but its non-trivial coupling to matter prevents singularities in early cosmology and in the non-relativistic collapse of non-interacting particles.
We extend our previous analysis, discussing further developments. We present a full numerical study of spherically symmetric non-relativistic gravitational collapse of dust. For any positive coupling, the final state of the collapse is a regular  pressureless star rather than a singularity. We also argue that there is no Chandrasekhar limit for the mass of non-relativistic white dwarf in this theory. Finally, we extend our previous results in the fully relativistic theory by constructing static and slowly rotating compact stars governed by nuclear-physics inspired equations of state. In the relativistic theory, there exists an upper bound on the mass of compact objects, suggesting that black holes can still be formed in the relativistic collapse.
\end{abstract}


\maketitle
\section{Introduction}\label{sec:intro}
The beauty and the beast of Einstein's General Relativity (GR) are encoded in the non-linearity of its field equations. Already in vacuum, GR describes the dynamics of non-linear objects, like black holes~\cite{Penrose:1996}. In order to study the formation of black holes in dynamical situations, e.g. during a stellar collapse, one needs to couple the vacuum theory to matter. How to include this coupling in a proper way was one of Einstein's main concerns. The requirement of stress-energy tensor conservation, $\nabla_\mu T^{\mu\nu}=0$, which in turn implies geodesics motion, together with Bianchi's identities, $\nabla_\mu G^{\mu\nu}=0$, naturally suggests a linear coupling between the Einstein tensor and the stress-energy tensor, $G_{\mu\nu}\propto T_{\mu\nu}$. 
Indeed, under quite generic assumptions (see e.g.~\cite{DInverno}) Einstein's equations are the most general field equations which involve the stress-energy tensor \emph{linearly}.
Nonetheless, it comes as a surprise that a highly non-linear theory as GR is just linearly coupled to matter. In this paper we investigate modified theories of gravity which account for a non-linear coupling to matter, while retaining many appealing features of Einstein's theory.

In the weak-field regime, GR has brilliantly passed all experimental and observational scrutinies so far~\cite{Will:2005va,Everitt:2011hp}, but new effects are still viable in the strong-field, non-linear regime. Furthermore, while current experiments have confirmed the weak-field behavior of GR \emph{in vacuum} or in orbital motion, performing null tests of Einstein's theory \emph{inside matter} may be extremely challenging because, due to the equivalence principle, purely gravitational effects are hard to disentangle from those due to non-standard matter coupling. In fact, one usually \emph{assumes} a minimal coupling, then solves the full Einstein equations in some relevant situation --~e.g. in cosmological settings, in the interior of Sun-like stars or inside more compact objects like neutron stars (NSs)~-- and finally compares theoretical models with observations.

On the other hand, GR suffers from severe theoretical problems and long-standing observational puzzles, which may be precisely related to our poor understanding of the gravity-matter coupling. For example, the dark matter problem (see e.g. Ref.~\cite{Bertone:2004pz} for a review) may be explained by invoking new fundamental interactions~\cite{Bekenstein:1984tv,Milgrom:1983ca}, rather than assuming exotic particles. Similarly, the cosmological acceleration of the universe may be explained in terms of more complicated interactions, rather than postulating the existence of a mysterious form of dark energy (see e.g. Ref.~\cite{Clifton:2011jh} for a recent review on cosmology in modified gravities). These postulates are somehow the modern versions of the aether, suggesting that perhaps something is missing in our understanding of gravitational interactions inside matter.

Likewise, it is well known that the dynamical evolution of matter fields in GR  is generically plagued by the formation of singularities (e.g. the Big Bang or those forming in the gravitational collapse of matter fields), which signal a break-down of the theory. According to Penrose's cosmic censorship~\cite{Penrose}, these singularities must be covered by an event horizon, i.e. a black hole must form as an outcome of the gravitational collapse. However, the cosmic censorship remains a conjecture and its validity (or even its precise formulation) is an open issue.

In a spherically symmetric stellar collapse, locally naked singularities can be generically formed~\cite{Lemos:1991uz,PhysRevLett.66.994,Joshi:1993zg}. Initial density gradients can produce shear in the collapsing fluid which, in turn, delays the formation of an apparent horizon, leaving the singularity locally naked~\cite{Joshi:2001xi}. In some cases, the singularity can even be globally naked, i.e. connected to distant observers by timelike or null-like geodesics, thus suggesting that the censorship can be evaded (see Refs.~\cite{Joshi:2000fk,Gundlach:2002sx,Joshi:2008zz} for detailed discussions on this topic). However, these ``shell-focusing'' singularities may be limited to the spherically symmetric case~\cite{Wald:1997wa}.
Relaxing the assumption of spherical symmetry, the situation is even more controversial. While Shapiro and Teukolsky reported numerical evidence for formation of naked singularities in the collapse of highly prolated gas spheroids~\cite{Shapiro:1991zza}, Wald and Iyer~\cite{PhysRevD.44.R3719} pointed out that  similar properties --~namely the absence of trapped surfaces lying on their maximal slices in a portion of singular spacetime~-- are also present in a Schwarzschild spacetime with a particular choice of the time slice. 
These studies show that, although there appears to be growing evidence in support of the cosmic censorship, its validity in GR remains an open issue which is far to be solved~\cite{Wald:1997wa,Lehner:2001wq}.

However, we stress here that the issue of singularity formation and the related cosmic censorship strongly depend on how gravity interacts with matter. Thus, the outcome of a gravitational collapse may be different if the gravity-matter coupling is modified.

In this paper, following previous works~\cite{Vollick:2003qp,Vollick:2005gc,Vollick:2006qd,Kerner:1982,Banados:2010ix,Pani:2011mg}, we take a different perspective and investigate phenomenologically viable modifications to GR that account for non-trivial coupling to matter while being equivalent to Einstein's theory in vacuum. As a consequence, this class of corrections results in a qualitatively different phenomenology once matter is included (see also Ref.~\cite{Bertolami:2007gv}). 
For concreteness, we will focus on a recent proposal by Ba\~nados and Ferreira~\cite{Banados:2010ix} (see also some previous attempt by Vollick~\cite{Vollick:2003qp,Vollick:2005gc,Vollick:2006qd}) which we shall call Eddington-inspired Born-Infeld~(EiBI) gravity, not to be confused with Eddington-Born-Infeld~\cite{Banados:2008rm} or Born-Infeld-Einstein theory~\cite{Deser:1998rj}. While being conceptually different from each other, these theories share the same determinantal form of Born-Infeld non-linear electrodynamics~\cite{Born:1934gh,Tseytlin:1999dj}. In this sense, similarly to non-linear electrodynamics, EiBI gravity can be thought as an effective prototypical theory in which resummation of higher order curvature terms --~which are qualitatively similar to those expected by a quantum gravity completion~-- is advocated in order to resolve curvature singularities~\cite{Klebanov:2000hb}. It turns out that these corrections effectively account for some non-linear matter coupling to usual Einstein's gravity.

Interestingly enough, such corrections can even affect the non-relativistic gravitational regime. We shall discuss the non-relativistic phenomenology in detail and most of our results apply to \emph{any} theory whose non-relativistic limit is described by the following modified Poisson equation~\cite{Banados:2010ix}
\begin{equation}
 \nabla^2\Phi=4\pi G\rho+\frac{\kappa}{4}\nabla^2\rho\,,\label{Poisson}
\end{equation}
where $\Phi$ is the usual gravitational potential, $G$ is the gravitational constant and $\rho$ is the matter density.
Within the parametrized post-Poissonian approach proposed in Ref.~\cite{Casanellas:2011kf}, the equation above is the most general spacially covariant Poisson equation, which is first order in $\Phi$ and $\rho$ and reduces identically to  standard Laplace's equation, $\nabla^2\Phi=0$, in vacuum. While possible second order terms are strongly constrained by tests of the equivalence between gravitational and inertial mass to one part in $10^{12}$(see e.g. Ref.~\cite{Su:1994gu} and references therein), the term proportional to $\kappa$ in Eq.~\eqref{Poisson} is presently mild constrained.

To our knowledge, null tests of Poisson equation inside matter, in order to constrain the extra term in Eq.~\eqref{Poisson}, have not been conceived yet. Mild observational constraints on $\kappa$ come from solar physics observations~\cite{Casanellas:2011kf}, $|\kappa|<3\times10^5 \text{m}^5 \text{s}^2 / \text{kg}$. This bound has been derived by comparing the observed solar neutrino fluxes and helioseismology observables to the predictions of modified solar models. Such constraints are quite mild, due to some inherent uncertainties on the Sun interior. Very recently, considerably stronger observational constraints were derived in Ref.~\cite{Avelino}, provided direct measures of the central density or of the radius of a NS are available (see also Ref.~\cite{Pani:2011mg}). However, some uncertainty exists on the interior of compact stars and usually GR is assumed to infer the magnitude of the central density, given some ``observable'' quantity such as the mass and the radius. Thus, table experiments in some controlled setting would be highly desirable.

The scope of this paper is twofold. First, we wish to discuss the phenomenology of Eq.~\eqref{Poisson} in relation with stellar collapse and static stellar configurations in the non-relativistic limit. Our purpose is to show that, while being compatible with current observations, a simple modified Poisson model leads to a new interesting phenomenology, even for arbitrarily small values of $\kappa$. Secondly, we shall discuss a well-motivated relativistic completion of Eq.~\eqref{Poisson}, EiBI gravity~\cite{Banados:2010ix}. In Ref.~\cite{Pani:2011mg}, we reported some results on the gravitational collapse and compact star solutions in this theory. Here, we give further details and discuss new developments. 

The plan of this work is the following. In Sec.~\ref{sec:theory} we briefly review some known and new features of EiBI gravity~\cite{Banados:2010ix}. Sections~\ref{sec:newtonian_collapse} and \ref{sec:nonrel_stars} are devoted to the phenomenology of the modified Poisson equation~\eqref{Poisson}, which includes the non-relativistic limit of EiBI gravity. Section~\ref{sec:newtonian_collapse} is devoted to solve the hydrodynamics equations describing the non-relativistic collapse of non-interacting particles. In pass, we shall correct a wrong result in Ref.~\cite{Pani:2011mg} which, however, does not affect the final picture. In Sec.~\ref{sec:nonrel_stars} we show that the end-point of the gravitational dust collapse is a regular, pressureless star and we discuss other non-relativistic stellar equilibrium configurations. Finally, in Sec.~\eqref{sec:rel_stars} we discuss relativistic stellar models using nuclear-physics based equations of state. In Sec.~\ref{sec:conclusion} we draw our conclusion and discuss open issues and future developments.
\section{Eddington-inspired Born-Infeld gravity}\label{sec:theory}
EiBI gravity is described by the following action~\cite{Banados:2010ix}
\be
S_g = \frac{2}{\kappa}\int d^4x\left(\sqrt{|g_{ab} + \kappa R_{ab}|} - \lambda \sqrt{-g}\right) +S_M\left[g,\Psi_M\right]\,,\label{action}
\ee
where $S_M\left[g,\Psi_M\right]$ is the matter action, $\Psi_M$ generically denotes any matter field, $R_{ab}$ is the symmetric part of the Ricci tensor built from the connection $\Gamma_{ab}^c$ and $\lambda$ is related to the cosmological constant, $\Lambda=(\lambda-1)/\kappa$, so that asymptotically flat solutions are obtained when $\lambda=1$.

EiBI gravity is reminiscent of a Born-Infeld action for non-linear electrodynamics~\cite{Born:1934gh,Tseytlin:1999dj}, where here the Ricci tensor plays the role of the field strength $F_{ab}$. Moreover, the metric $g$ and the connection $\Gamma$ are considered as independent fields~\cite{Banados:2010ix}, as in Palatini's approach to GR. Similarly to $f(R)$ gravities (see e.g.~\cite{Sotiriou:2008rp}) the Palatini formulation is not equivalent to the metric one. In the metric approach, EiBI theory contains ghosts, which must be eliminated by adding extra terms to the action~\cite{Deser:1998rj,Vollick:2003qp}. Furthermore, it is assumed that matter minimally couples to the metric tensor only, i.e. the matter action $S_M$ in Eq.~\eqref{action} only depends on the metric $g$ and on the matter fields, but not on the independent connection $\Gamma$.

The field equations are conveniently written as~\cite{Banados:2010ix}
\begin{eqnarray}
 \Gamma_{ab}^c(q)&=&\frac{1}{2}q^{cd}\left(\partial_a q_{bd}+\partial_b q_{ad}-\partial_d q_{ab}\right)\,,\label{Gamma}\\
 q_{ab}&=&g_{ab}+\kappa R_{ab}(\Gamma)\,,\label{eqDIN}\\
 \sqrt{-q}q^{ab}&=&\lambda\sqrt{-g}\left(g^{ab}-\kappa T^{ab}\right) \label{eqALG}
\end{eqnarray}
where $q_{ab}$ is an auxiliary metric, $T^{ab}$ is the standard stress-energy tensor and $R_{ab}(\Gamma)$ is the symmetric part of the Ricci curvature tensor. Crucially, since matter is minimally coupled to the metric $g_{ab}$, the field equations above imply the usual conservation of the stress-energy tensor, $\nabla_a T^{ab}=0$, where $\nabla$ is the covariant derivative written in terms of the metric tensor $g$. Note that $q_{ab}$ is the matrix inverse of $q^{ab}$ and differs inside matter from $q^{cd}g_{ac}g_{bd}$.

When $T^{ab}\equiv0$, the auxiliary metric $q$ coincides with the physical metric $g$. Thus, in vacuum $\Gamma^a_{bc}$ is the metric connection. Indeed, in absence of matter fields EiBI theory is equivalent to GR~\cite{Banados:2010ix}. In presence of matter, the small $\kappa$ limit of the field equations reads
\begin{equation}
R_{ab}(\Gamma)=T_{ab}-\frac{1}{2}Tg_{ab}+\kappa\left[S_{ab}-\frac{1}{4}Sg_{ab}\right]+{\cal O}(\kappa^2)\,.\label{eqs_small_kappa}
\end{equation}
where $S_{ab}={T^c}_a T_{c b}-\frac{1}{2}T T_{ab}$.
Notice that Einstein's theory is recovered as $\kappa\to0$. In the field equation above, two qualitatively different corrections to GR are manifest at ${\cal O}(\kappa)$: those depending on $S_{ab}$, which are quadratic in the matter fields, and those hidden in $R_{ab}(\Gamma)$, which implicitly depend on derivatives of matter fields. The latter survive even in the non-relativist limit of the theory, which is described by a modified Poisson equation~\eqref{Poisson}.

Since the auxiliary metric $q_{ab}$ is algebraically related to the physical metric $g_{ab}$, EiBI gravity does not introduce any extra dynamical degree of freedom with respect to GR. In this sense, the theory is similar in spirit to $f(R)$ gravities in the Palatini approach~\cite{Sotiriou:2008rp}.

\subsection{Two formulations of EiBI gravity}

Let us now investigate the field equations of EiBI theory in more detail. Written in terms of the new rank-two auxiliary tensor $q$, the theory is reminiscent of a particular bi-metric theory (see e.g. Refs.~\cite{Drummond:2001rj,Milgrom:2009gv} for some specific attempt) which, in vacuum, degenerates into a single metric theory, GR. On the other hand, inside matter the two metrics are different and the connection is metric with respect to $q$, and not $g$, cf. Eq.~\eqref{Gamma}. Nonetheless, the conservation of the stress-energy tensor --~and geodesic motion therein~-- is guarantee by a minimal gravity-matter coupling in the action~\eqref{action}. This is precisely the essence of the theory.

It should be noted however that there are at least two ways of interpreting the theory to which we shall refer to, with some abuse, as the ``Jordan frame'' and the ``Einstein frame'', in accordance to scalar-tensor theories. In the Jordan frame matter is minimally coupled to the metric $g$ and Einstein equations are modified according to Eq.~\eqref{eqs_small_kappa}. In the Einstein frame the metric $q$ is treated as fundamental and matter is non-minimally coupled to it. This appears quite naturally in the small $\kappa$ limit of the theory, Eq.~\eqref{eqs_small_kappa}, where $g_{\mu\nu}$ can be written solely in terms of $q_{\mu\nu}$ and its derivatives by using Eq.~\eqref{eqDIN}.

Let us however start by pointing a particular symmetry of the action. In the Jordan frame, the action reads as in Eq.~\eqref{action}. In what we define as the Einstein frame, the action can be written explicitly by defining a new metric $\gamma_{ab}=g_{ab}+\kappa R_{ab}(\Gamma)$,
\bea
S_\gamma &=& \frac{- 2\lambda}{\kappa}\int d^4x\left( \sqrt{|\gamma_{ab} - \kappa R_{ab}(\Gamma)|}-\frac{1}{\lambda}\sqrt{-\gamma} \right)\nonumber\\ 
&&+S_M\left[\gamma_{ab} - \kappa R_{ab}(\Gamma),\Psi_M\right]\,.\label{action_gamma}
\eea
Written in this form, the metric $\gamma$ is now non-minimally coupled to matter. Furthermore, the field equations impose $\gamma_{ab}=q_{ab}$, but the connection $\Gamma$ is not anymore the $q$-metric connection, since $\Gamma$ enters explicitly in the matter action through $R_{ab}(\Gamma)$. 

Neglecting the matter action in $S_g$ and $\ S_\gamma$, the following duality holds
\be
g_{ab}\leftrightarrow \gamma_{ab}\,,\quad \kappa\leftrightarrow -\kappa\,,\quad  \lambda \leftrightarrow 1/\lambda\,, \quad \Gamma\leftrightarrow\Gamma\,,\label{duality}
\ee
up to an overall factor $\lambda$. This duality can be seen in the vacuum field equations, where the rescaling follows from the algebraic relation \eqref{eqALG}. Note that only in the case $\lambda=1$, i.e. for a vanishing cosmological constant,  $\gamma_{ab} = g_{ab}$. This also shows that the case $\lambda=1$ is special not only because it corresponds to asymptotically flat space in both frames, but it is a fixed point of the duality~\eqref{duality}. 
If $\lambda=1$, negative values of  $\kappa$ in the Einstein frame correspond to positive values of $\kappa$ in the Jordan frame, provided one modifies the matter coupling. 

However, the coupling to matter breaks the symmetry above, introducing a non-minimal coupling in the Einstein frame. 
Interestingly, the action~\eqref{action_gamma} in the Einstein frame reduces, in the small $\kappa$ limit, to Einstein's theory with non-minimally coupled matter (see also Ref.~\cite{Bertolami:2007gv}), 
\bea
S_\gamma &=& { \lambda}\int d^4x\sqrt{-\gamma} \left( R - 2 \frac{\lambda-1}{\kappa\lambda} \right) \nonumber\\
&&+ \int d^4x\sqrt{-\gamma}\left(1- \frac{1}{2}\kappa R(\Gamma)\right) \mathcal L_m(\gamma_{ab}-\kappa R_{ab}(\Gamma))\nn\\
&& + {\cal O}(\kappa^2) \,.
\label{linSq}
\eea
where we have defined $R=\gamma^{ab}R_{ab}$. 

Furthermore, if the matter Lagrangian is constructed in terms of the metric $g$, the second term in Eq.~\eqref{linSq} involves other $\kappa$ corrections and some higher order coupling to the Ricci tensor and to the independent connection.
For example, let us consider the small $\kappa$ limit of a minimally coupled free scalar field in the Jordan frame
\be
\mathcal L_m(g_{ab},\phi) = g_{ab}\partial^a\phi\partial^b\phi\,.
\ee
In the Einstein frame, the matter Lagrangian reads
\be
\mathcal L_m(\gamma_{ab},\Gamma,\phi) = \gamma_{ab}\partial^a\phi\partial^b\phi - \kappa R_{ab}(\Gamma)\partial^a\phi\partial^b\phi\,,
\ee
leading to the following matter action in the Einstein frame
\bea
S_m&=&\int d^4x\sqrt{-\gamma}\Big[\gamma_{ab}\partial^a\phi\partial^b\phi\\
&&-\kappa \left( R_{ab}+\frac{1}{2} R \gamma_{ab}\right)\partial^a\phi\partial^b\phi + {\cal O}(\kappa^2)\,.\nn
\eea
Note that the action above involves second derivatives of the scalar field in the gravity equations hidden inside the Ricci tensor evaluated on shell. 

A similar transformation as the one presented above persists in the non-relativistic limit. Indeed, if we define $\phi=\Phi+\frac{\kappa}{4}\rho$, Eq.~\eqref{Poisson} reduces to the standard Poisson equation,
\be
\nabla^2\phi=4\pi G\rho\,.\label{Poisson_q}
\ee
However, in this case the gravitational force would read $F=\nabla\Phi=\nabla\phi-\frac{\kappa}{4}\nabla\rho$, being thus dependent on the matter field.

The modified Poisson equation~\eqref{Poisson} has an interesting interpretation in terms of standard Poisson equation with a modified source term. The second term on the right hand side of Eq.~\eqref{Poisson} leads to an effective force:
\begin{equation}
 \frac{F_{\rm eff}}{\rho}\equiv-\frac{\nabla P_{\rm eff}}{\rho}=-\frac{\kappa}{4}\nabla\rho\,, \label{Peff}
\end{equation}
from which we obtain that $P_{\rm eff}=\kappa \rho^2/8$. Therefore, the modified Poisson equation is formally equivalent to the standard one supplemented by an effective polytropic fluid with equation of state (EOS) $P(\rho)=K\rho^{1+1/n}$, where the polytropic index $n=1$ and $K=\kappa/8$. In fact, a similar property exists also in the relativistic case. In the weak field limit of EiBI theory, one can show that an effective pressure and effective internal energy appears. This will be explained in detail in a forthcoming work~\cite{TerenceJan}.

\subsection{Linear structure of EiBI theory}
The linear dynamics of EiBI theory is equivalent to linearized Einstein's theory. To prove this, we expand the field equations~\eqref{eqDIN}-\eqref{eqALG} around a vacuum Minkowski background,
\begin{eqnarray}
&& q_{ab}=\eta_{ab}+\delta q_{ab}\,,\qquad  g_{ab}=\eta_{ab}+\delta g_{ab}\,,\\
&& q^{ab}=\eta^{ab}-\delta q^{ab}\,,\qquad  g^{ab}=\eta^{ab}-\delta g^{ab}\,,\\
&& T_{ab}=\delta T_{ab}\,,\qquad  T^{ab}=\delta T^{ab}\,,
\end{eqnarray}
where indeces are raised and lowered by the Minkowski metric, $\eta_{ab}$. In order for the Minkowski metric to be solution, we set $\lambda=1$ in the field equations.
Linearization of Eq.~\eqref{eqDIN} follows immediatly from the standard linearized Ricci tensor,
\begin{equation}
 \delta q_{ab}-\delta g_{ab}=\kappa\delta R_{ab}\equiv\frac{\kappa}{2}\left(\delta q^c_{a,bc}+\delta q^c_{b,ac}-\square\delta q_{ab}-\delta q_{ab}\right)\,,\label{eqDINlin}
\end{equation}
where a coma denotes partial derivative with respect to the flat background and $\delta q$ is the trace of $\delta q_{ab}$.
In order to linearize Eq.~\eqref{eqALG}, we recall that, at linear order, $|-g|=1+\delta g$, so we get
\begin{equation}
\delta q^{ab}-\frac{\eta^{ab}}{2}(\delta q-\delta g)=\delta g^{ab}+\kappa \delta T^{ab}\,\label{eqALGlin}
\end{equation}
By taking the trace of the equation above, we get $\delta q-\delta g=-\kappa \delta T$, which can be substituted back into Eq.~\eqref{eqALGlin}. Finally Eq.~\eqref{eqDINlin} can be written as
\begin{equation}
 \delta R_{ab}=\delta T_{ab}-\frac{\eta_{ab}}{2}\delta T\,,\label{lin_fin}
\end{equation}
which does not depend on $\kappa$.
This equation has exactly the same form as in GR, but the Ricci tensor is written in terms of $q_{ab}$. However, at linear order, we note that
\begin{equation}
 \delta T_{ab}=\left.\frac{\delta T_{ab}}{\delta\Psi_M}\right|_{\rm vacuum}\delta\Psi_M+\left.\frac{\delta T_{ab}}{\delta g^{lm}}\right|_{\rm vacuum}\delta g_{lm}\,,
\end{equation}
where the second term vanishes when evaluated in vacuum, since matter fields are coupled to the metric. Therefore, $\delta T_{ab}$ and $\delta T$ do not depend explicitly on the metric and Eq.~\eqref{lin_fin} is exactly equivalent to the linearized Einstein equations to \emph{all orders} in $\kappa$, but for the auxiliary metric~$q_{ab}$.

In particular, the linearized version of EiBI theory admits the same differential structure as linearized GR. For example the principal symbols, i.e. terms in the linear equations involving highest order derivatives, are the same as GR. 
Although a detailed analysis of the well-posedness of the theory is beyond our scope, these results suggest that there exists a well-posed formulation of EiBI theory, similarly to GR. 
For a unitarity analysis of general BI gravity theories see Ref.~\cite{Gullu:2010em}.

Finally, in the small $\kappa$ limit, our initial assumption of Minkowski background can be relaxed. In this case, similar conclusion about the linear structure of EiBI around any background metric can be drawn.

\section{Non-relativistic stellar collapse}\label{sec:newtonian_collapse}
In this section, we describe the non-relativistic collapse of dust in EiBI theory. Hereafter, we focus on asymptotically flat solutions, setting $\lambda=1$. 
The collapse of incoherent dust in the Newtonian limit shares many properties with its relativistic analogue~\cite{Oppenheimer:1939,Florides1977138}. 
Here, we shall solve the hydrodynamics equations in the case of spherical symmetry, showing that the end-state of the $1+1$ evolution is the pressureless star found in Ref.~\cite{Pani:2011mg} and described in detail in Section~\ref{sec:nonrel_stars} below.
This confirms and extends the analytical argument presented in Ref.~\cite{Pani:2011mg}. However, at the end of this section we show that our analytical computation in Ref.~\cite{Pani:2011mg} is partially flawed, due to a typo in the series expansion close to the origin, and we shall discuss why the numerical results are still in agreement with the overall picture given in Ref.~\cite{Pani:2011mg}.

The hydrodynamical equations can be equivalently solved using either the Eulerian or the Lagrangian approach. In the next two sections, we shall briefly review these formulations, together with a standard procedure to avoid shock wave formation, and the  modifications related to the problem at hand.
\subsection{Eulerian formulation}

In the non-relativistic limit, the collapse of non-interacting ($P=0$) particles is governed by the mass conservation (continuity equation) and momentum conservation (Euler equation), the latter being modified in EiBI gravity due to Eq.~\eqref{Poisson}.

Following the seminal work~\cite{vonneumann:232}, we supplement the hydrodynamics equations by an artificial viscosity term, in order to avoid divergences due to shock wave formation during the evolution. In the Eulerian formulation, the relevant equations in the pressureless case read~\cite{Florides1977138,Iwakami:2007ie} 
\begin{eqnarray}
\frac{\partial\rho}{\partial t}+\mathbf{u}\cdot\nabla\rho+\rho \nabla\cdot\mathbf{u}&=&0\,,\label{Eul2}\\
\frac{\partial\mathbf{u}}{\partial t}+\mathbf{u}\cdot\nabla \mathbf{u}+\nabla\Phi+\frac{\nabla\cdot\mathbf{Q}}{\rho}&=&0\,,\label{euler}\\
m-\int d^3x \rho&=&0\,,\label{Eul1}
\end{eqnarray}
where $\mathbf{u}$ is the fluid velocity, $\rho$ is the density and $m$ is the mass function. Note that the second equation above is vectorial, $\mathbf{Q}$ is the viscosity tensor and $\nabla\cdot\mathbf{Q}$ is a vector.

The viscosity tensor is defined as~\cite{Iwakami:2007ie}
\begin{equation}
 \mathbf{Q}=\ell^2\rho \nabla\cdot\mathbf{u}\left[\nabla\mathbf{u}-\frac{\mathbf{e}}{3}\nabla\cdot\mathbf{u} \right]
\end{equation}
if $\nabla\cdot\mathbf{u}<0$, otherwise $\mathbf{Q}=0$.
In the equation above $\nabla\mathbf{u}=(\partial_j u_i+\partial_i u_j)/2$ and $\mathbf{e}$ is the unit tensor. Note that this equation only contains scalar invariant quantities so that can be directly specialized to the spherically symmetric case, contrarily to Eq.~(8) in Ref.~\cite{vonneumann:232}, which is only valid in Cartesian coordinates.

In indicial form,
\begin{equation}
 Q_{ij}=\ell^2\rho\nabla\cdot\mathbf{u}\left[\frac{\partial_j u_i+\partial_i u_j}{2}-\frac{\delta_{ij}}{3}\nabla\cdot\mathbf{u} \right]\,,
\end{equation}
%
where $\ell$ is a constant with dimension of length. 

We are  interested in the spherical symmetry case, where all the dynamical variable only depends on $(t,r)$ and the only non-vanishing component of vectorial quantities above is on the radial direction, e.g. $\mathbf{u}=(u(t,r),0,0)$. In this case only the radial component of Eq.~\eqref{euler} is nontrivial and $\left[\nabla\cdot\mathbf{Q}\right]_r=\nabla\cdot\mathbf{q}$, where $q_i=Q_{ir}$. Due to the symmetry, the vector $\mathbf{q}=(q(t,r),0,0)$, and
\begin{equation}
 q(t,r)=\ell^2 \rho(t,r)\nabla\cdot\mathbf{u}\left[u'(t,r)-\frac{1}{3}\nabla\cdot\mathbf{u}\right]\,.
\end{equation}
Finally, in spherical symmetry, $\nabla\cdot\mathbf{u}=u'+2u/r$ and the artificial viscosity term in Eq.~\eqref{euler} is $\nabla\cdot \mathbf{q}=q'+2q/r$, so that we are left with the set of PDEs
\begin{eqnarray}
\frac{\partial\rho}{\partial t}+u\rho'+\rho u'+\frac{2}{r}\rho u&=&0\,,\\
\frac{\partial u}{\partial t}+u u'+\frac{G m}{r^2}+\frac{\kappa}{4}\rho'+\frac{{\cal Q}}{\rho}&=&0\,,\label{eulerF}\\
m'-4\pi r^2\rho&=&0\,,
\end{eqnarray}
where a dot and a prime denote derivatives with respect to $t$ and $r$, respectively. The viscosity term reads
\begin{eqnarray}
 {\cal Q}&=&\frac{2 \ell^2}{3 r^2} \left[r \rho\left(u+2 r u'\right) u''\right.\nn\\
&&\left.-\left(u-r u'\right) \left(3 \rho u'+\left(2 u+r u'\right) \rho'\right)\right]\,,
\end{eqnarray}
if $u'+2u/r<0$, otherwise  ${\cal Q}=0$. In practice, $\ell=c \Delta r$, where $\Delta r$ is the grid spacing in the radial direction and $c$ is some dimensionless constant. This allows the shock to be ``smeared'' in a region of width $\ell$, while leaving the rest of the dynamics unchanged. In our simulations, we have checked that the results do not depend on the precise value of $c$. Actually, the results do not even depend on the details of the artificial viscosity term, as long as it introduces a smearing effect on the shocks while keeping the rest of the physics unaltered. We have explicitly checked this fact, by comparing test simulations with different artificial terms.  

Finally, one can derive a Bernoulli-like relation by multiplying Eq.~\eqref{euler} by $u(t,r)$, leading to 
\be
\frac{d}{dt}\left( \frac{1}{2}u(t,r)^2 +\Phi\right) = \frac{\partial \Phi}{\partial t}\,.\label{energycons}
\ee
where we neglected the artificial viscosity term and where $d/dt$ is the total time derivative. This expression is valid locally and the term on the right hand side takes into account the local variation due to matter flow. Indeed in the pure Newtonian case, this term can be rewritten as a surface integral of the density current.

\subsection{Lagrangian formulation}
The basic Lagrangian equations are given in \cite{vonneumann:232}. They are expressed in terms of the comoving volume 
$V(t,x)= \frac{1}{\rho_0}\nabla X(t,x)$ and the velocity field $u=\frac{\partial X(t,x)}{\partial t}$, where $\rho_0(x)$ is the initial
density. The continuity equation for the volume is given by
\be
\rho_0(x)\dot V(t,x) = \dot X(t,x),
\ee
which, in terms of the density $\rho(t,x)=1/V(t,x)$, reads
\be
\dot \rho = -\rho \frac{u'(t,x)}{X'(t,x)} = -\rho \frac{\p u(t,x)}{\p X(t,x)}.
\ee

If $X(t,x)=R(t,x)$ is a radial variable in spherical coordinates, this equation becomes
\be
\dot\rho(t,x) + \frac{2}{R(t,x)}\rho(t,x) u + \frac{\rho(t,x) u'(t,x)}{R'(t,x)} = 0.
\ee

Note that the equation above can be simply obtained from the corresponding Eulerian formulation by replacing the Eulerian time derivative by the Lagrangian time derivative according to
\be
\frac{D}{Dt} = \frac{\partial }{\partial t} + \mathbf{u}\cdot\mathbf{\nabla},
\ee
where $D/Dt$ is the time derivative in the Lagrangian formulation.

Following the same principle, the equation for the velocity field is given by
\be
\dot u = -\frac{G m(t,x)}{R(t,x)^2} - \kappa \frac{\rho'(t,x)}{R'(t,x)} + \frac{F_q(t,x)}{\rho_0(x)},
\ee
where $F_q(t,x)$ is the artificial viscosity term given by~\cite{vonneumann:232} which explicitly reads
\be
F_q = \frac{\partial}{\partial x}\left(\frac{\rho_0(x) (c \Delta x)^2 u'(t,x)|u'(t,x)|}{R'(t,x)}\right),
\ee
where $\Delta x$ is the grid spacing and $c$ is a constant.
\subsection{Results}\label{collapse_results}
We have solved the initial-value problem defined by the above system of PDEs using the method of lines, in which the radial dimension is discretized and the resulting system of coupled ODEs is integrated in time with standard methods. The initial static profiles  we considered are presented in Table~\ref{tab:initprof}.

\begin{table}[!h]
 \begin{tabular}{c|cc}
		&	$\rho(0,r)$ & $u(0,r)$  \\
\hline
Profile I	&	$e^{-\sigma r^2}$ & $0$ \\
Profile II	&	$\frac{\sin\varpi r}{\varpi r}e^{-\sigma r^2}$ & $0$
\end{tabular}
\caption{Initial density and velocity profiles of our simulations.}
\label{tab:initprof}
\end{table}
Profile~I is a classical exponential density profile, while Profile~II is a deformation of the pressureless star found in Ref.~\cite{Pani:2011mg} (cf. Eq.~\eqref{pressureless_Newtonian} below). For both class of profiles, $\sigma$ is a free parameter related to the slope of the initial density profile at the center.

For testing purposes, we wrote two independent codes that solve the hydrodynamical equations in the Eulerian and in the Lagrangian formulation, respectively. Once convergence is reached, the results of the two codes agree very well. However, the simulations in the Lagrangian formulation generically need a lower resolution, due to the fact that the radial grid evolves along with the fluid collapse. Indeed, the Lagrangian formulation is analog to the comoving frame generically adopted in relativistic collapse simulations. Hence, we shall present results obtained using this formulation. Typically, we used a non-uniform grid with $4\times10^3$ points in the spacial direction, which guarantees convergence of the results for all values of $\kappa$ taken into account. As expected, we find that, for a given value of the viscosity constant $c$, convergence and stability of the numerical results is reached provided the mesh is sufficiently fine.

In standard Newtonian gravity, $\kappa=0$, the hydrodynamics equations can be solved analytically for a constant density profile and they correspond to the relativistic Oppenheimer-Snyder solution~\cite{Florides1977138}. In that case, the dust collapses in a finite time $t_C=\pi\sqrt{R^3/(8M)}$, where $R$ and $M$ are the initial radius and the total mass of the spherical dust configuration. This analytical results is also valid when non-constant initial density profiles are considered~\cite{Pani:2011mg}.
Our simulations reproduce this result with very good precision. Furthermore, for any negative value of $\kappa$, we found a qualitatively similar behavior, i.e. the fluid collapses to a singular state in a finite time. For $\kappa<0$, the time of collapse decreases with $|\kappa|$.

\begin{figure*}[htb]
\begin{center}
\begin{tabular}{cc}
\epsfig{file=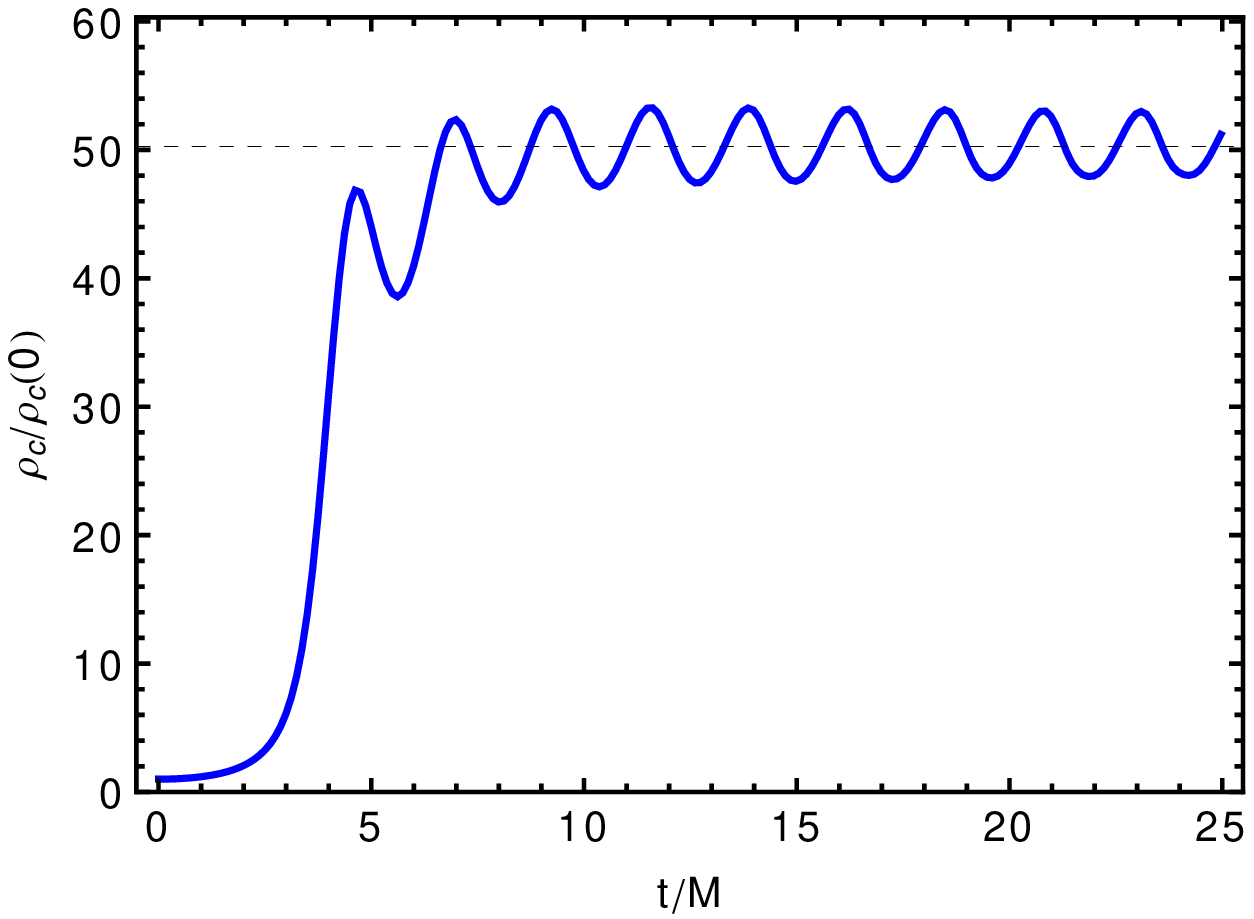,width=6.5cm,angle=0}&
\epsfig{file=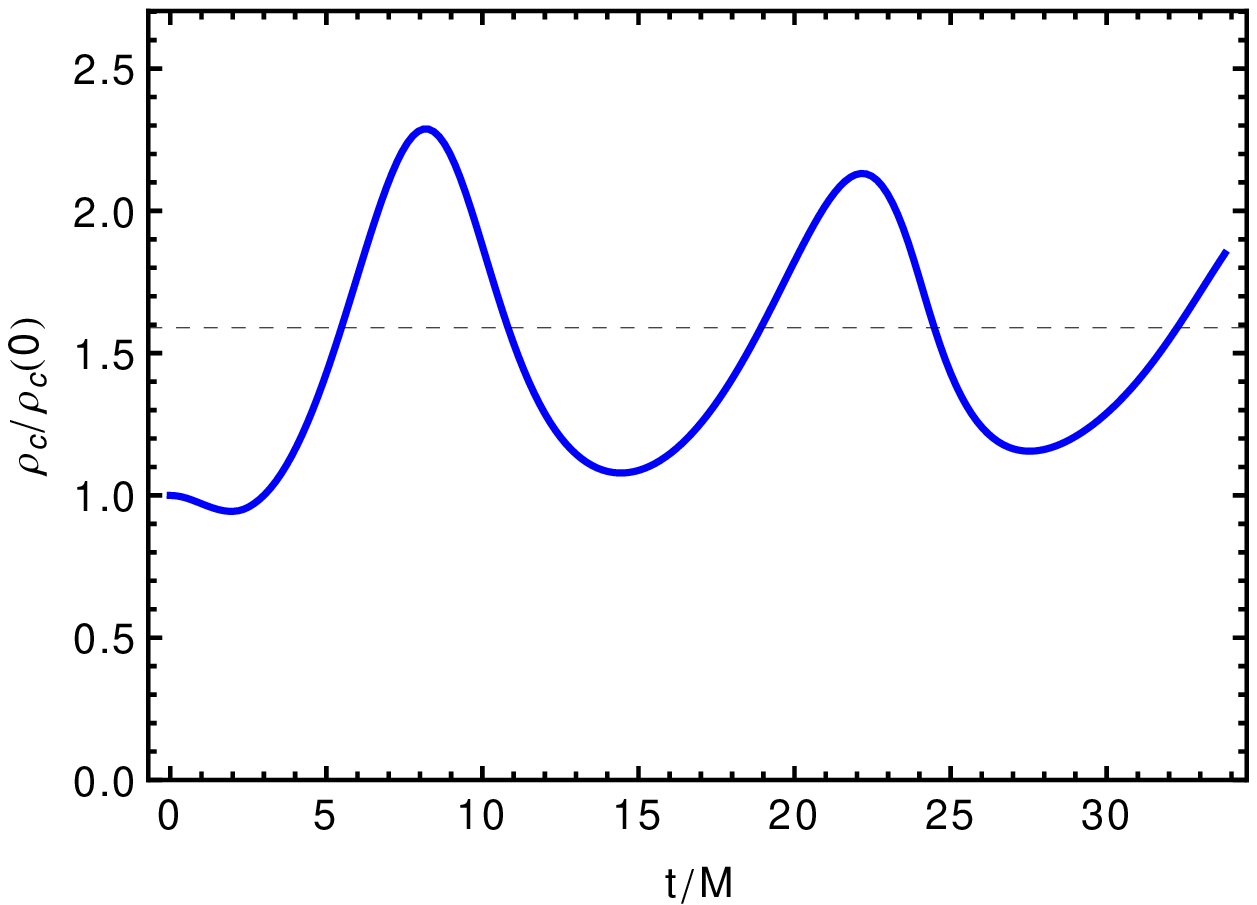,width=6.5cm,angle=0}
\end{tabular}
\caption{Central density as a function of time normalized by its initial value. Left panel: $\kappa\rho_c=0.1$. Right panel: $\kappa\rho_c=1$. Dashed lines denotes the central density of the pressureless star with the same mass (cf. Sec.~\ref{sec:pressureless}). The oscillation period agrees very well with the fundamental period of proper oscillation of these solutions (cf. Sec.~\ref{sec:stability} and Fig.~\eqref{fig:period})
\label{fig:collapse}}
\end{center}
\end{figure*}

\begin{figure*}[htb]
\begin{center}
\begin{tabular}{cc}
\epsfig{file=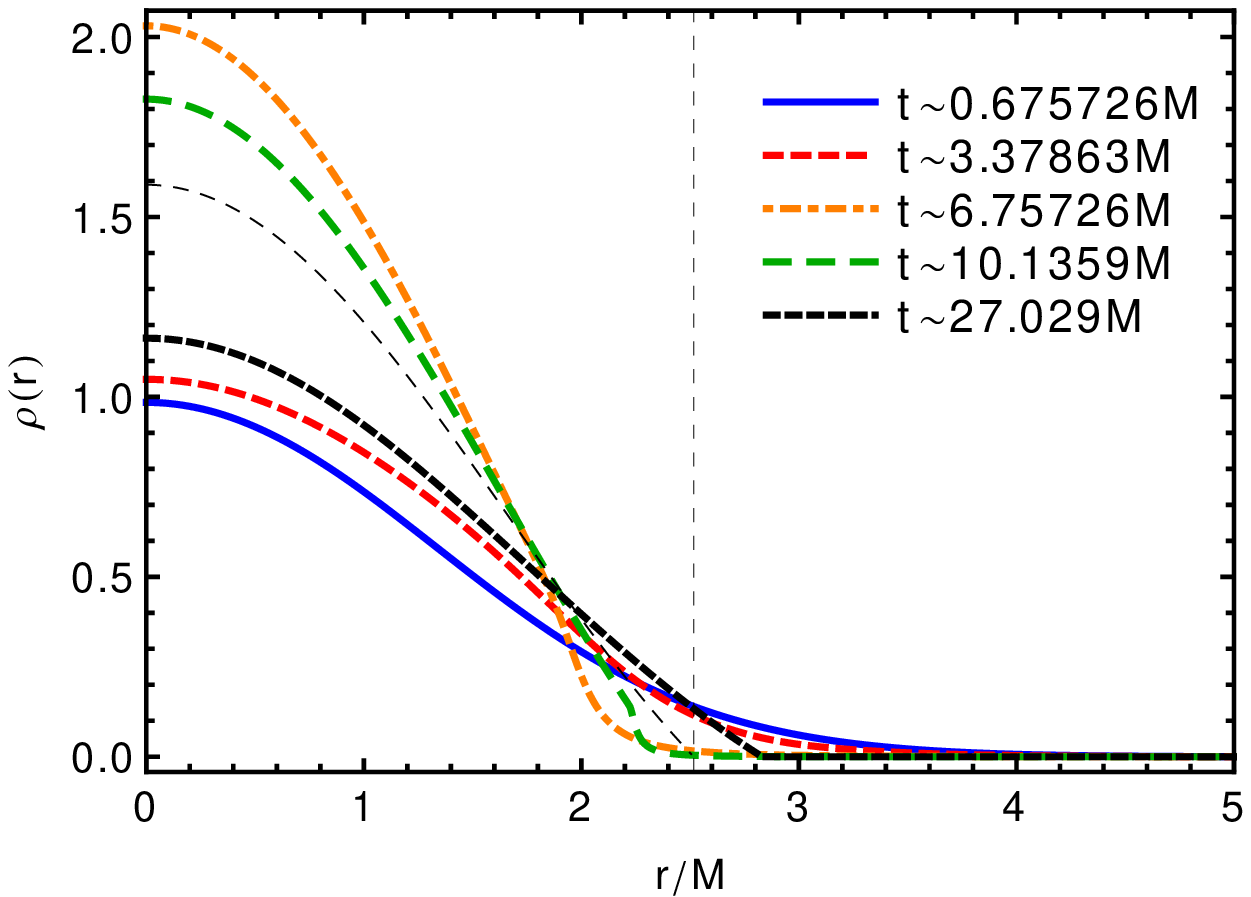,width=6.5cm,angle=0}&
\epsfig{file=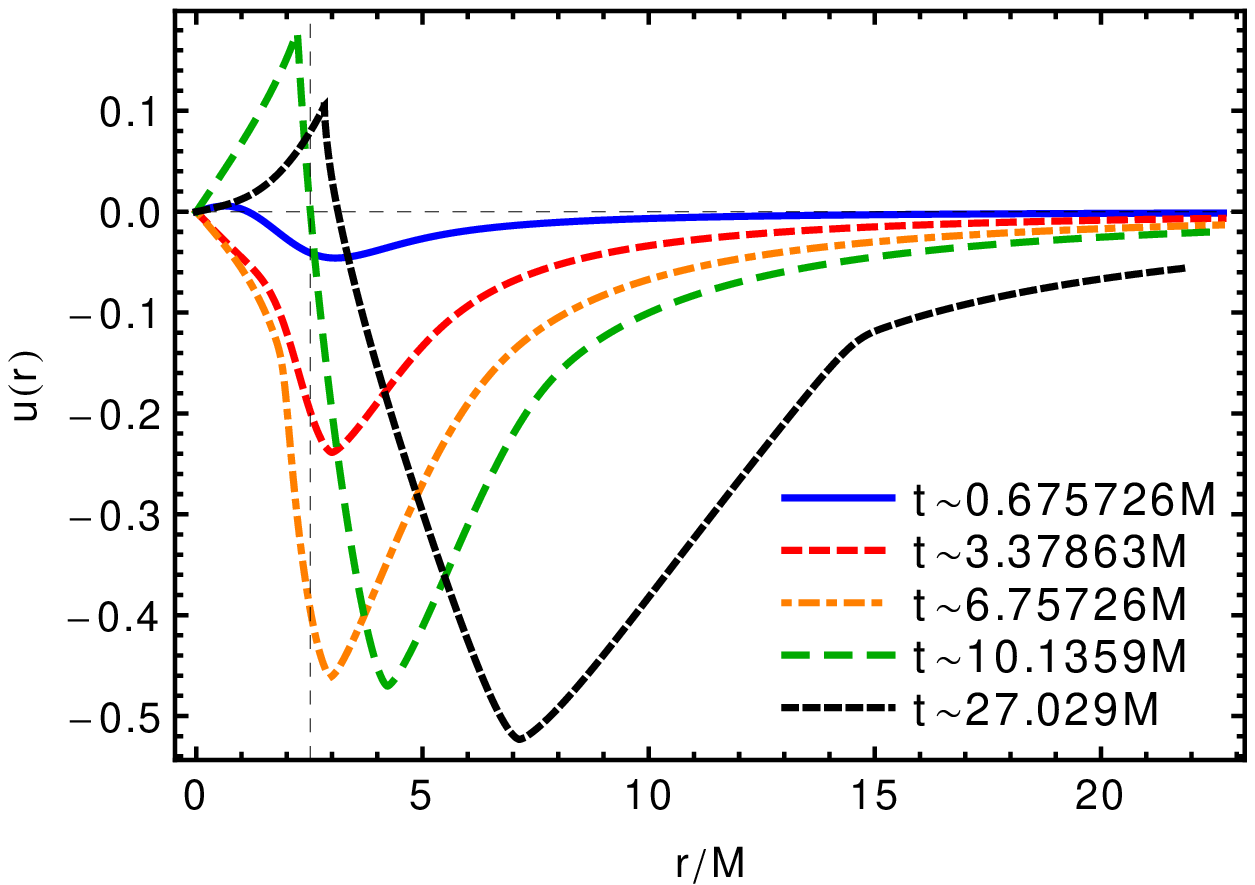,width=6.5cm,angle=0}
\end{tabular}
\caption{Radial profiles of the density (left panel) and of the velocity (right panel) for different instants and for $\kappa\rho_c=1$. The dashed vertical line denotes the radius of the pressureless star (cf. Sec.~\eqref{sec:pressureless}) with the same mass and the dashed black thin line in the left panel denotes the pressureless star profile.
\label{fig:profiles}}
\end{center}
\end{figure*}

On the other hand, the behavior for any $\kappa>0$ is drastically different, as shown in Figs.~\ref{fig:collapse} and \ref{fig:profiles}. We present results for $\kappa\rho_c=0.1,1$ and obtained using Profile I in Table~\eqref{tab:initprof}, but different choices of $\kappa>0$, different initial profiles or different values of $\sigma$, would give qualitatively similar results.
In Fig.~\ref{fig:collapse} we show the central density as a function of time for two different values of $\kappa$. The central density is always finite and oscillates around a constant value at late time. As we shall show in Sec.~\eqref{sec:pressureless}, the mean value of the oscillations (denoted by an horizontal dashed line in Fig.~\ref{fig:collapse}) is precisely the central density of a pressureless star~\cite{Pani:2011mg} (cf. Sec.~\eqref{sec:pressureless}) with the same mass. In Fig.~\ref{fig:profiles}, we show the density and the velocity radial profiles at different instants. The would-be shocks in the velocity profile (the steep region shown in the right panel of Fig.~\ref{fig:profiles}) propagates towards the exterior of the fluid without developing a discontinuity, which could not be resolved due to finite grid spacing. The artificial term discussed above smears this discontinuity and ensures stability of the numerical simulations.

\begin{figure}
 \begin{center}
\epsfig{file=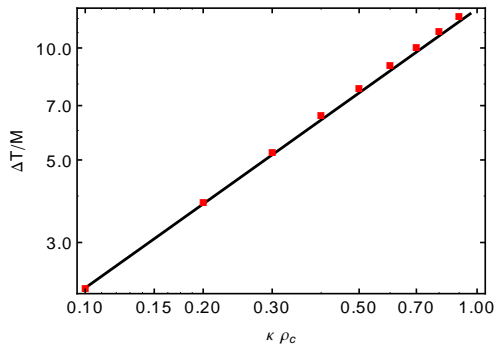,width=6.5cm,angle=0}  
 \end{center}
\caption{Period of the oscillations around the pressureless static configuration as a function of $\kappa$ and with gaussian initial profile ($\sigma=10$). The straight line is the prediction from the linear stability analysis (cf. Eq.~\eqref{omemass}) whereas the markers are extracted from the numerical simulations.}
\label{fig:period}
\end{figure}

Our simulations suggest that, even for an arbitrarily small value of $\kappa$, this oscillatory behavior would continue indefinitely. This is perfectly consistent with the fact that the hydrodynamical equations do not include any dissipative term, so that energy is conserved during the evolution. Nevertheless, our results give strong evidences that, for generic (static) initial profiles, the end-point of the dust collapse is precisely the pressureless star. Indeed, we expect that any dissipative term, which has to be included in realistic situations, would quench the oscillations and drift the system toward a static configuration, which is precisely the pressureless star. This picture is also confirmed by Fig.~\ref{fig:period}, which shows the period of oscillation, as a function of $\kappa$, compared with the fundamental period of proper oscillation of a pressureless star (cf. Sec.~\ref{sec:stability}).

Finally, as shown in Eq.~\eqref{Peff}, the pressureless collapse in the modified non-relativistic theory is equivalent to the collapse of a polytropic fluid with $P(\rho)=\kappa\rho^2/8$ in Newtonian theory. Hence, our results are consistent with the fact that, in the latter case, the final state is a regular polytropic star with polytropic index $n=1$ (see e.g.~\cite{Larson:1969mx} for more realistic simulations of star formation in Newtonian gravity).
\subsubsection{On the analytical method of Ref.~\cite{Pani:2011mg}}
In Ref.~\cite{Pani:2011mg}, we developed an approximated method to solve the collapse equation analytically for any $\kappa$. Here, we point out a typo in that computation and we discuss why this does not affect the final result, as we have proved in the previous section by solving the collapse equation numerically. 

Let us review the procedure adopted in Ref.~\cite{Pani:2011mg}. Due to the spherical symmetry, we expect the singularity to initially form at the center $r=0$. The collapse equations are then solved by expanding the dynamical variables close to the center,
\begin{eqnarray}
 \rho&=&\rho_0(t)+\rho_1(t)r+\rho_2(t) r^2+{\cal O}(r^3)\,,\nn\\
 u&=&u_0(t)+u_1(t)r+u_2(t)r^2+u_3(t)r^3+{\cal O}(r^4)\nn\,,
\end{eqnarray}
and solving a system of ODEs for $\rho_i(t)$ and $u_i(t)$. At first order, one finds $\rho_1(t)=u_0(t)=u_2(t)=0$ and
\begin{equation}
 u_1(t)=-\frac{\dot\rho_0(t)}{3\rho_0(t)}\,.
\end{equation}
At second order, we get
\begin{eqnarray}
 \frac{\ddot\rho_0(t)}{\rho_0(t)}- \frac{4}{3} \frac{\dot\rho_0(t)^2}{\rho_0(t)^2}-\frac{3}{2}\kappa\rho_2(t)-8\pi G\rho_0(t) &=&0\,\label{eqcollapse1}\\
 \frac{\dot\rho_2(t)}{\rho_2(t)}- \frac{5}{3} \frac{\dot\rho_0(t)}{\rho_0(t)}+5 \frac{\rho_0(t)}{\rho_2(t)}u_3(t)&=&0 \,.\label{eqcollapse2}\
\end{eqnarray}
The results we presented in Ref.~\cite{Pani:2011mg} originated by erroneously omitting the term proportional to $u_3(t)$ in Eq.~\eqref{eqcollapse2}. Indeed, if $u_3(t)\equiv0$, then $\rho_2(t)\propto\rho_0^{5/3}(t)$, which can be substituted in Eq.~\eqref{eqcollapse1} to obtain a single non-linear ODE for $\rho_0(t)$. The solution of that equation is given in the equation below Eq.~(7) of Ref.~\cite{Pani:2011mg}. 

However, in general $u_3(t)$ is non-vanishing and Eqs.~\eqref{eqcollapse1} and \eqref{eqcollapse2} are not sufficient to solve for the three variables $u_3$, $\rho_0$ and $\rho_2$. In fact, it can be easily proved that, at any order in the series expansion, the number of unknown functions is always larger than the number of differential equations. This is precisely due to the extra derivative term in the Poisson equation~\eqref{Poisson}. Indeed, in the standard case when $\kappa=0$, Eq.~\eqref{eqcollapse1} decouples and can be solved in the usual way. As a matter of fact, when $\kappa\neq0$, the collapse equations cannot be solved exactly by this simple method, as erroneously stated in Ref.~\cite{Pani:2011mg}.

Nevertheless, as we have showed above, our fully numerical simulations not only confirm the results previously reported, i.e. that the collapse does not lead to any singularity, but also that the final state is the pressureless solution reported in Ref.~\cite{Pani:2011mg}. Furthermore, the oscillatory behavior found in the numerical simulation qualitatively match the oscillatory behavior presented in Fig.~1 of Ref.~\cite{Pani:2011mg}. 

Thus, it is relevant to understand whether the analytical method present above, although flawed, can actually reproduce the full solution in some limit. This is certainly true when ${\rho_0(t)}u_3(t)\ll{\rho_2(t)}$ at any time.
The latter condition is indeed satisfied when the velocity and density profiles are close enough to a late-time evolving configuration, for example when the configuration oscillates around the static configuration. Indeed in this case, $\rho_0(t) = \rho_0^s + {\cal O}(\epsilon),\ \rho_2 =\rho_2^s + {\cal O}(\epsilon)$ while $u_3(t)={\cal O}(\epsilon)$, where $\rho_0^s,\rho_2^s$ are the constant central density and second derivative at the center of the static  pressureless configuration and $\epsilon$ is a small number. It follows that in this case, the combination $\rho_0 u_3$ is one order smaller than $\rho_2$.

\section{Stars in the non-relativistic limit}\label{sec:nonrel_stars}
In this section, we discuss non-relativistic stellar models in the theory defined by~\eqref{action}. Our results apply to any relativistic theory which reduces to Eq.~\eqref{Poisson} in the non-relativistic regime.
We shall solve the hydrostatic equilibrium equation, supplemented by the standard mass conservation, $dm/dr=4\pi r^2\rho(r)$, and an EOS. Requiring spherical symmetry, the hydrostatic equilibrium equation reads~\cite{Pani:2011mg}
\be
\frac{dP}{dr}=-\frac{Gm(r)\rho}{r^2}-\frac{\kappa}{4}\rho\rho'\,.\label{hydroeq2}
\ee
Clearly, equilibrium configurations in EiBI gravity are different from the standard Newtonian ones in presence of non-trivial density profiles. Note that the equation above can be written in a more evocative form as~\cite{Casanellas:2011kf}
\begin{equation}
\frac{dP}{dr}=-G_\text{eff}(r)\frac{m(r)\rho(r)}{r^2}\,,\label{hydroeq_eff}
\end{equation}
where we have defined an ``effective'' Newton's constant
\begin{equation}
 G_\text{eff}(r)\equiv G+\frac{\kappa}{4}\frac{r^2\rho'(r)}{m(r)}\,.\label{Geff}
\end{equation}
Since $\rho'(r)<0$ inside a star, $G_\text{eff}\lessgtr G$ when $\kappa\gtrless0$. In Ref.~\cite{Casanellas:2011kf}, the modified hydrostatic equilibrium equation has been used to construct realistic solar models and put the first observational constraint on $\kappa$ (see also the recent Ref.~\cite{Avelino} for more stringent constraints).

Here, we shall mainly focus on polytropic models with EOS of the form $P=P(\rho)$. Using the mass conservation equation, we can rewrite Eq.~\eqref{hydroeq2} as
\begin{equation}
 \frac{1}{r^2}\left[\frac{r^2}{\rho}\rho'\frac{dP(\rho)}{d\rho}\right]'+4\pi  G \rho+\frac{\kappa }{4}\left(\rho''+2 \frac{\rho'}{r}\right)=0\,,\label{eqpoly}
\end{equation}
where a prime denotes derivative with respect to $r$.
Once the EOS is fixed, Eq.~\eqref{eqpoly} is a second order ODE for $\rho(r)$, which can be solved by imposing regularity condition at the center, i.e.
\be
\rho(r)=\sum_{n=0}^\infty\rho_{n}r^{n}\,, \quad r\to0\,.
\ee
In the equation above, $\rho_i$ are constants which can all be expressed in terms of the only free parameter, $\rho_0$, by solving Eq.~\eqref{eqpoly} perturbatively close to the center. For example, 
\begin{equation}
 \rho_2=-\frac{8 G \pi  \rho_0^2}{3 (4 P'(\rho_0)+\kappa  \rho_0)}\,,\label{rho2}
\end{equation}
whereas $\rho_l=0$ for any odd $l$.
It is easy to see that stellar models only exist when $\rho_2<0$, which implies a lower bound, $\kappa\rho_0>-4 P'(\rho)$. As we shall see, this limit holds also qualitatively in the relativistic limit and in more realistic models~\cite{Casanellas:2011kf}. 
\subsection{Non-relativistic pressureless stars}\label{sec:pressureless}
In classical Newtonian gravity, a gas of non-interacting particle cannot support self-graviting configurations and it will inevitably collapse due to its self-gravity attraction. However, in theories in which the Poisson equation reads as in Eq.~\eqref{Poisson}, a straightforward solution of the hydrostatic equilibrium equation with $P\equiv0$ and $\kappa>0$ reads~\cite{Pani:2011mg}
\begin{equation}
 \rho(r)={\rho_c}\frac{\sin(\varpi r)}{\varpi r}\,,\qquad \varpi=4\sqrt{\frac{G\pi}{\kappa}}\,.\label{pressureless_Newtonian}
\end{equation}
The radius and mass of the star read
\begin{equation}
 R=\frac{\pi}{\varpi}\,,\qquad M=\frac{4\pi^2\rho_c}{\varpi^3}\,,
\end{equation}
respectively, so that for a given $\kappa>0$, the solution above is uniquely characterized by the central density or, equivalently, by the mass. In particular, the radius of the compact object is uniquely determined by $\kappa$.

The existence of such solutions is remarkable not only because they are interesting models for self-gravitating dark matter~\cite{Bertone:2004pz}, but also because they provide a non-trivial static and spherically symmetric solution of the hydrodynamics equations governing the collapse of dust. Indeed, as we discussed in Sect.~\ref{sec:newtonian_collapse}, our numerical simulation give strong evidences that, as an outcome of the collapse, any initial distribution of matter will eventually relax towards the pressureless solution described by Eq.~\eqref{pressureless_Newtonian}. 

Clearly, the generality of this result is only due to the fact that we considered the collapse of pressureless matter, but it also suggests that, when taking into account realistic models in which the matter is described by a specific EOS, the end-state of the collapse will be a modified compact star, rather than a singular state. A detailed analysis in that direction would be certainly interesting. For the time being, it is important to notice that these solutions are also linearly stable, as first shown in Ref.~\cite{Pani:2011mg} and detailed in Sec.~\ref{sec:stability}.

We note here that the pressureless star~\eqref{pressureless_Newtonian} is a sort of solitonic solution, since it solves the Poisson equation for any harmonic function $\Phi$, due to the identically vanishing of the right hand side of Eq.~\eqref{Poisson}, i.e. $G_{\rm eff}=0$. In the interior, the Newtonian potential is constant and it matches continuously the vacuum potential $M/r$ at the radius. 
Indeed, Eq.~\eqref{Poisson} can be thought as a forced oscillator equation for $\rho$, where $\kappa$ would play the role of the inverse spring constant and $\Phi$ is an external force. The pressureless star qualitatively corresponds to the solution of the oscillator with a constant external force.

\subsection{Newtonian polytropic models}
For a generic polytropic index $n$, the field equation must be solved numerically, imposing $\rho\sim\rho_0+\rho_2 r^2$ at the center. In this case stellar solutions only exist provided the following condition is satisfied
\be
\kappa>-|\kappa_c|=-4 K (1+1/n) \rho_c^{-1+1/n}\,.\label{kcrit}
\ee
For $\kappa>0$, condition (\ref{kcrit}) is always fulfilled. 
In some cases the Lane-Emden equation obtained from Eq.~\eqref{hydroeq2} can be solved analytically~\cite{Chandra_Book_Stars}. For instance
if $n=1$, $P(\rho)=K\rho^2$, and the solution reads as in~\eqref{pressureless_Newtonian}, but with $\varpi=4\sqrt{G\pi/(8K+\kappa)}$, so that
it exists for $\kappa>-8K$ and reduces to the pressureless case for $K=0$. This is consistent with the effective pressure term defined in~\eqref{Peff}. Indeed, the modified pressureless star solution~\eqref{pressureless_Newtonian} with $\kappa=8K$ is exactly the same as the $n=1$ polytropic solution with $\kappa=0$.

\subsection{Modified Chandrasekhar model for white dwarfs}
\begin{figure}[htb]
\begin{center}
\begin{tabular}{c}
\epsfig{file=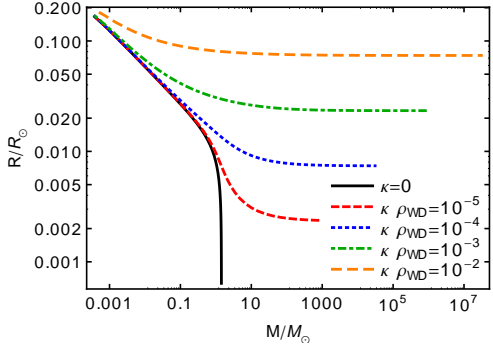,width=6.5cm,angle=0}
\end{tabular}
\caption{Mass-radius relation for zero temperature white dwarfs for different values of $\kappa$ (in units of a typical white dwarf density, $\rho_{WD}=10^9\text{kg/m}^3$) in EiBI theory and compared to the standard result for $\kappa=0$. Notice that, for any $\kappa>0$, there is no maximum mass, but white dwarf models have a minimum radius given by Eq.~\eqref{wdcrit}.
\label{fig:Chandra}}
\end{center}
\end{figure}

As a relevant example of non-relativistic polytropic star, let us consider zero temperature non-rotating white dwarfs. 
The interior of a zero-temperature white dwarf is described by a relativistic EOS, $P(\rho)$, which is parametrically defined by~\cite{Chandrasekhar:1935zz}
\begin{eqnarray}
 P(x)&=& \frac{\pi m_e^4 c^5 \mu_e m_P}{3 h^3}\left[x(2x^2-3)\sqrt{1+x^2}+3\sinh^{-1}x \right]\,,\nn\\
 \rho(x)&=& \frac{8\pi m_e^3 c^3 \mu_e m_P}{3 h^3}x^3\,,\label{EOS_WD}
\end{eqnarray}
where $c$ and $h$ are the speed of light and the Planck constant, $m_e$ and $m_P$ are the electron and the proton mass, respectively and $\mu_e$ is the molecular weight. In Fig.~\ref{fig:Chandra} we show the mass-radius relation for different values of $\kappa>0$ obtained by integrating numerically the hydrostatic equations.
%

When $\kappa=0$ we recover the famous Chandrasekhar result: zero temperature white dwarfs have a maximum mass, $M\approx 1.4 M_\odot$. However, for any $\kappa>0$, the behavior is drastically different. White dwarf models can have an arbitrarily large mass, but instead there exists a minimum radius, which approximately scales as $\kappa^{1/2}$.

We can use Landau's original argument to understand these results. Following Shapiro and Teukolsky's exposition~\cite{Shapiro:1983du}, let us consider a star of radius $R$ composed of $N$ fermions, each of mass $m_B$. In the relativistic regime, the Fermi energy of the system reads
\be
E_F=\frac{\hbar cN^{1/3}}{R}\,.
\ee
In EiBI theory, the gravitational energy per fermion is approximately
\be
E_G \approx -\frac{GNm_b^2}{R}+\frac{3\kappa Nm_b^2}{16\pi R^3}\,,
\ee
where we approximated the star's density as $\rho\sim Nm_b/(4\pi R^3/3)$.
Thus, the star's total energy read
\be
E \equiv E_F+E_G= \frac{\hbar cN^{1/3}-GNm_b^2}{R}+\frac{3\kappa Nm_b^2}{16\pi R^3}\,,
\ee
For small $N$, the total energy is positive, and we can decrease it by increasing $R$.
If moreover 
\be
R>R_{\rm crit}\equiv \sqrt{\frac{3\kappa}{16\pi G}}\,,\label{wdcrit}
\ee
then the total energy can become negative. At some point the electrons become non-relativistic and the purely Newtonian term dominates. It is easy to see that when this happens the energy tends to zero when $R$ goes to infinity. Thus, there is a point of minimum energy, and it is delimited by the $R=R_{\rm crit}$ curve \eqref{wdcrit}. Note that the scaling $R_{\rm crit}\sim\sqrt{\kappa}$ is consistent with the recent computation of the Jeans length in Ref.~\cite{Avelino}.

On the other hand, even if $E<0$ initially, for sufficiently small $R$ condition (\ref{wdcrit}) will be violated: the EiBI term dominates at sufficiently small radii, and collapse is never energetically favored.
Relativistic stellar models of white dwarfs are described in Sec.~\ref{sec:WDrel}.

\subsection{Stability in the non-relativistic limit}\label{sec:stability}
We now discuss stability of the Newtonian configurations against {\it radial} perturbations~\cite{Shapiro:1983du}, expanding the discussion in Ref.~\cite{Pani:2011mg}. The equations governing the stellar dynamics are the Poisson equation~\eqref{Poisson} together with the continuity equation and the momentum equation. These are given in Eqs.~\eqref{Eul2}-\eqref{Eul1} by replacing the artificial viscosity term by a physical pressure term, i.e. $\nabla\cdot\mathbf{Q}\to \nabla P$. In those equations, the hydrostatic equilibrium is recovered when $\mathbf{u}=0$.

We shall focus on spherically symmetric models.
The perturbed Euler equation reads
\begin{equation}
 \Delta\left[\frac{du}{dt}+\frac{P'}{\rho}+\Phi'\right]=0\,.\label{DeltaMomentum}
\end{equation}
where $d/dt=\partial/\partial t+\mathbf{u}\cdot\nabla$ is the total time derivative, $\Delta=\delta+\xi\cdot\nabla$ and $\xi$ is the Lagrangian displacement.
In order to compute this equation explicitly we use the following~\cite{Shapiro:1983du}
\begin{eqnarray}
 \Delta\rho&=&-\frac{\rho}{r^2}(r^2\xi)'\,,\nn\\
 \delta\rho&=&-\frac{(r^2\rho\xi)'}{r^2}\,,\nn\\
 (\delta\Phi)'&=&-4\pi\rho\xi+\frac{\kappa}{4}(\delta\rho)'=-4\pi\rho\xi-\frac{\kappa}{4}\left[\frac{(r^2\rho \xi)'}{r^2}\right]'\,,\nn\\
 \Delta P&\equiv& P\gamma\frac{\Delta\rho}{\rho}\nn\,,
\end{eqnarray}
where the last equation above defines the adiabatic index of the perturbations, $\gamma$. It is then straightforward to obtain the following
\begin{eqnarray}
 \Delta\left(\frac{du}{dt}\right)&=&\frac{d^2\xi}{dt^2}\,,\nn\\
 \Delta\left(\frac{P'}{\rho}\right)&=&-\frac{\Delta\rho}{\rho^2}P'+\frac{\Delta(P')}{\rho}=\frac{2\xi}{r\rho}P'+\frac{(\Delta P)'}{\rho}\nn\\
 &=&\frac{2\xi}{r\rho}P'-\frac{1}{\rho}\left[P\gamma\frac{(r^2\xi)'}{r^2}\right]'\,,\nn\\
 \Delta(\Phi')&=&(\Delta\Phi)'-\xi'\Phi'=(\delta\Phi)'+\xi\Phi''\nn\\
 &=&(\delta\Phi)'+\xi\nabla^2\Phi-\frac{2}{r}\xi\Phi'\nn\\
 &=&\frac{2\xi}{\rho r}P'+\frac{\kappa}{4}\left[\frac{2}{r}\xi\rho'-\xi'\rho'-\left[\frac{\rho}{r^2}(r^2\xi)'\right]'\right]\,,\nn
\end{eqnarray}
where we have used $\Delta{d}/{dr}={d}/{dr}\Delta-\xi'{d}/{dr}$ and the background equations.
Finally, using the equations above, Eq.~\eqref{DeltaMomentum} reads
\begin{eqnarray}
 &&\ddot\xi-\frac{1}{\rho}\left[\frac{\gamma P}{r^2}(r^2\xi)'\right]'+\frac{4}{\rho r}\xi P'\nn\\
&&+\frac{\kappa}{4}\left[\frac{2}{r}\xi\rho'-\xi'\rho'-\left[\frac{\rho}{r^2}(r^2\xi)'\right]'\right]=0\,.
\end{eqnarray}
Assuming a time dependence $e^{i\omega t}$, the modified eigenvalue equation reads~\cite{Pani:2011mg}
\be
\!\!\frac{4\xi P'}{\,r} \!+\!\frac{\kappa\rho}{4}\left[\frac{2}{r}\xi\rho'\!\!-\!\!\xi'\rho'\!\!-\!\!\left[\frac{\rho}{r^2}(r^2\xi)'\right]'\right]\!\!-\!\!\left[\frac{\gamma P}{r^2}(r^2\xi)'\right]'\!\!=\rho\xi\omega^2\,.\label{eigeinvalue_eq}
\ee
This equation must be solved for $\xi$, requiring the following boundary conditions~\cite{Shapiro:1983du}
\begin{equation}
 \xi(0)=0\,,\qquad \Delta P(R)=-\gamma\frac{P}{r^2}(r^2\xi)'=0\,,\label{BCB}
\end{equation}
the latter being equivalent to requiring regularity of $\xi$ at the radius. An instability corresponds to an eigenmode with $\omega^2<0$.

\subsubsection{Perturbations of pressureless stars}
For $P\equiv0$ and $\rho$ given by Eq.~\eqref{pressureless_Newtonian}, the eigenvalue equation~\eqref{eigeinvalue_eq} simplifies. In particular it does not depend on the index of perturbations $\gamma$, but only on $\kappa$. Thus, for a given $\kappa$, there is one fundamental mode. By integrating Eq.~\eqref{eigeinvalue_eq} numerically and imposing Eqs.~\eqref{BCB}, we find
\begin{equation}
 \omega = \alpha\rho_c^{1/2}\,,
\end{equation}
where $\alpha \approx 2.1866$, independently from $\kappa$. 

This gives the characteristic period of oscillation 
 \begin{equation}
  \frac{\Delta T}{M}= \frac{\pi ^{5/4}}{2 \alpha } \left(\frac{\kappa }{M^2}\right)^{3/4}\,,
\label{omemass}
 \end{equation}
which is shown in Fig.~\ref{fig:period} and compared with the period of oscillation of our numerical solutions around a pressureless star with same mass. The results of the linear analysis perfectly agrees with the simulations, giving further support for the end-state of the collapse.

In addition, we found no unstable modes of the pressureless star, confirming that in the modified Newtonian theory these object are linearly stable~\cite{Pani:2011mg}.

\subsubsection{Newtonian polytropic stars}
\begin{figure}[htb]
\begin{center}
\begin{tabular}{c}
\epsfig{file=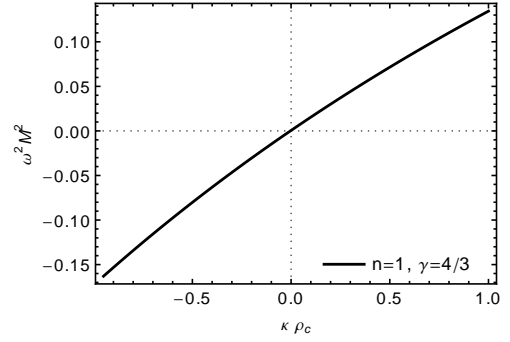,width=6.5cm,angle=0}
\end{tabular}
\caption{Fundamental mode of the polytropic star with $n=1$ and adiabatic index of perturbations $\gamma=4/3$. In the standard theory ($\kappa=0$) the star is marginally stable, whereas it is stable for $\kappa>0$ and unstable for $\kappa<0$.
\label{fig:fundamental_mode}}
\end{center}
\end{figure}

In Newtonian gravity, polytropic models with $\gamma=4/3$ are marginally stable for any polytropic index $n$~\cite{Shapiro:1983du}. In our case, these models are stable if $\kappa>0$ and unstable if $\kappa<0$. 
A representative example is shown in Fig.~\ref{fig:fundamental_mode}, where we show the fundamental mode as a function of $\kappa$ for the polytropic model with $n=1$ and $\gamma=4/3$.
For generic values of $\gamma$, positive values of $\kappa$ contribute to stabilize the models, while negative values work in the opposite direction. 
%

\section{Relativistic compact stars}\label{sec:rel_stars}
In this section, we study compact objects in the fully relativistic theory defined by the action~\eqref{action}. Let us now consider static and spherically symmetric perfect fluid stars. The metric ansatzen read
\bea
q_{ab}dx^a dx^b &=& -p(r) dt^2 + h(r) dr^2 + r^2 d\Omega^2,\label{ansatzq}\\
g_{ab}dx^a dx^b &=& -F(r) dt^2 + B(r) dr^2 + A(r)r^2 d\Omega^2\,,\label{ansatzg}
\eea
where we have used the gauge freedom to fix the function in front of the spherical part of the auxiliary metric $q$. 

Notice that the field equations~\eqref{eqALG} are simply algebraic equations relating $q$ and $g$. Inserting the ansatzen above into Eqs.~\eqref{eqALG}, and solving for the coefficients of $g$ in terms of $q$, leads to
\bea	
F &=& p(r) [1+\kappa  \rho]^2A^3(r),\label{alg1}\\
B &=& h(r)A(r),\label{alg2}\\
A &=& \left[(1 -\kappa  P)(1+\kappa  \rho)\right]^{-1/2}\,.\label{alg3}
\eea
which correctly imply $g_{\mu\nu}=q_{\mu\nu}$ in vacuum.
The dynamical field equations~\eqref{eqDIN} read
\bea
F&=&p-\frac{\kappa  \left(p' \left(r p h'+h \left(-4 p+r p'\right)\right)-2 r h p p''\right)}{4 r h^2 p} \,,\label{DIN1}\\
B&=&\frac{r F h^2-\kappa  \left(p h'+h p'\right)}{r h p}\,,\label{DIN2}\\
A&=&1-\frac{\kappa }{4r^2}\left(4+\frac{2 r h'}{h^2}-\frac{4}{h}-\frac{2 r p'}{hp}\right)\,,\label{DIN3}
\eea
where $F$, $B$ and $A$ are given in Eqs.~\eqref{alg1}-\eqref{alg3}. The equations above, supplied by an EOS in the form $P=P(\rho)$,  can be solved for $p$, $h$ and $P$. 
Since matter is covariantly coupled with the metric $g$, the standard conservation of the stress-energy tensor follows, $\nabla_a T^{ab}=0$, where we recall that the covariant derivative is defined in terms of the physical metric, $g_{ab}$. 
We consider perfect-fluid stars with energy density $\rho(r)$ and pressure $P(r)$ such that
\be
T^{ab}\equiv T^{ab}_{\rm perfect\,fluid}=\left[\rho+P\right]u^a\,u^b+g^{ab}P\,,\label{Tmunu_fluid}
\ee
where the fluid four-velocity $u^a=(1/\sqrt{F},0,0,0)$. Thus, rather than using Eq.~\eqref{DIN3}, it is more convenient to use the conservation of the stress-energy tensor,
\begin{equation}
 2 F P'+F'(P+\rho)=0 \,.\label{divT2}
\end{equation}
Notice that, although the equation above simply reads as in GR, $F'$ introduces terms proportional to $\rho'$ through Eq.~\eqref{alg1}. Finally, in the limit $\kappa\to0$, the field equations reduce to that of GR with $8\pi G=1$. 
\subsection{Integration}
We have integrated Eqs.~\eqref{DIN1}, \eqref{DIN2} and~\eqref{divT2} imposing regularity conditions at the center of the star, where the following expansions hold
\begin{eqnarray}
&&p(r)\sim p_0+p_2 r^2\,,\qquad h(r)\sim 1+h_2 r^2\,,\nn\\
&&P(r)\sim P_c+P_2 r^2\,,\qquad \rho(r)\sim \rho_c+\rho_2r^2\,.\nn
\end{eqnarray}
We can set $p_0=1$ by a time reparametrization. Furthermore, using the field equations, the coefficients $(p_2,h_2,P_2)$ can be written in terms of $P_c$ and $\rho_c$ as follows
\begin{eqnarray}
P_2&=&\frac{2 (1-P_c \kappa ) (P_c+\rho_c) (1+\kappa  \rho_c) \left({(1+\kappa  \rho_c)^2}A_c^3-1\right)}{3 \kappa\Delta}\,,\nn\\
\Delta&=&(P_c \kappa -3 \kappa  \rho_c-4) (1+\kappa  \rho_c)-\kappa  (1-\kappa P_c  ) (P_c+\rho_c) \rho'_c\,,\nn\\
p_2&=&\frac{{(1+\kappa  \rho_c)^2}A_c^{3}-1}{3 \kappa } \,,\quad h_2=\frac{2+{(1+\kappa  \rho_c)^2}A_c^{3}-{3}A_c}{6 \kappa }\nn \,,
\end{eqnarray}
with $\rho_c'={d\rho(P_c)}/{dP}$, $A_c=A(0)$ and where we have used $\rho_2=P_2 \rho_c'$. 

The series expansion of the field equations at the center of the star contains terms of the form
$A_c=\sqrt{(1-\kappa P_c)(1+\kappa \rho_c)}$. Assuming $\rho_c,P_c>0$, $\kappa$ must satisfy two conditions in order to allow for self-gravitating objects:
\begin{eqnarray}
&&P_c\kappa<1\,,\qquad\,\, \text{for $\kappa>0$} \,,\label{lim1}\\
&&\rho_c|\kappa|<1\,,\qquad \text{for $\kappa<0$}\,.\label{lim2}
\end{eqnarray}
Hence, assuming NSs may reach a central density $\rho_c\sim10^{18}$~kg\ m$^{-3}$ and $P_c\sim 10^{34}$~N\ m$^{-2}$, the bounds above strongly constrain the theory, $|\kappa|\lesssim 1\mbox{ m}^5\mbox{kg}^{-1}\mbox{s}^{-2}$ (see also the recent Ref.~\cite{Avelino}).
Furthermore, it is easy to prove that compact objects only exist if $P_2<0$, which requires $\kappa\Delta<0$. This gives a further constraint depending on $\rho_c$, $P_c$ and $\rho'_c$. The form of the constraint is cumbersome, but it is qualitatively equivalent to Eq.~\eqref{kcrit}. In particular, the condition is always satisfied for $\kappa>0$.

\subsection{Israel-Darmois matching conditions}

The field equations are integrated outward up to the radius $R$, defined by the condition $P(R)=0$, where we require the numerical solution to match the exact, and unique, vacuum Schwarzschild solution, $F(r)=B(r)^{-1}=p(r)=h(r)^{-1}=1-2 M/r$, where $M$ is the mass of the star. 

In GR, the mass is computed by imposing Darmois-Israel matching conditions~\cite{Israel:1966rt} at the radius, i.e. $[g_{ij}]=0$ and $[K_{ij}]=0$, where $[...]$ is the jump across the surface, $K_{ij}$ is the extrinsic curvature tensor, and $i,j=0,2,3$.
These matching conditions comes from the Einstein equations and the requirement of a well-defined $3-$geometry. For a spherically symmetric spacetimes, this leads to $[g_{tt}]=0=[g_{tt}']$ and $[g_{rr}]=0=[g_{rr}']$ through the field equations.

The same procedure can be applied to EiBI theory, as we explain in detail in Appendix~\ref{app:matching}. In this case, the junction conditions read
\begin{equation}
 [g_{ij}]=0\,,\quad [q_{ij}]=0\,,\quad [K_{ij}(q)]=\left[\frac{\partial_r q_{ij}}{\sqrt{q_{rr}}}\right]=0\,,\label{matchingBEI}
\end{equation}
where $K_{ij}(q)$ is the extrinsic curvature tensor for a hypersurface of a manifold $({\cal M},q)$. 
\begin{figure}[htb]
\begin{center}
\begin{tabular}{c}
\epsfig{file=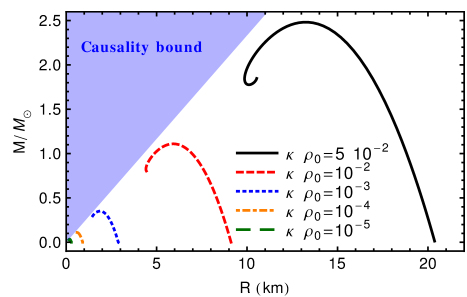,width=6cm,angle=0}
\end{tabular}
\caption{Mass-radius relation for a relativistic pressureless stars. Results are normalized by $\rho_0=8\cdot 10^{17}$~kg\ m$^{-3}$, which is a typical central density for NSs. The shaded region is excluded by causality, $R\gtrsim 2.9 GM$~\cite{Lattimer:2006xb} (cf. also the discussion at the end of Section~\ref{sec:realisticEOS}).
\label{fig:p0MR}}
\end{center}
\end{figure}
%
\subsection{Relativistic pressureless stars}\label{sec:relativistic_pressureless}
The existence of Newtonian pressureless stars makes it relevant to investigate the existence of similar solutions in the full theory. To this purpose, we set $P\equiv0$. The conservation of the stress-energy tensor simply implies $F(r)=$const. 
For a given value of $\kappa$, the solutions of the field equations then depend only on the central density.

As shown in Fig.~\ref{fig:p0MR}, for any value of $\kappa>0$, there exists a regular solution. These stars have a maximum compactness of $GM/R\sim0.3$, which is roughly independent from $\kappa$, a maximum mass\footnote{We thank Jan Steinhoff for having pointed out this property.} and a maximum radius which linearly increase with $\kappa$. Furthermore, they have a positive binding energy, $\bar{m}/M-1$, where $\bar{m}$ is the baryonic mass, corresponding to the
energy that the system would have if all baryons were dispersed to
infinity.

Of course, such compact objects do not exist in GR, while they exist in EiBI theory because $\kappa>0$ introduces a repulsive gravity contribution. Interestingly, the EOS for dark matter particles is approximately $P\equiv0$. Hence, in this theory self-gravitating objects, purely made by dark matter, can exist and may reach the typical compactness and mass of most compact NSs. 

\begin{figure*}[htb]
\begin{center}
\begin{tabular}{cc}
\epsfig{file=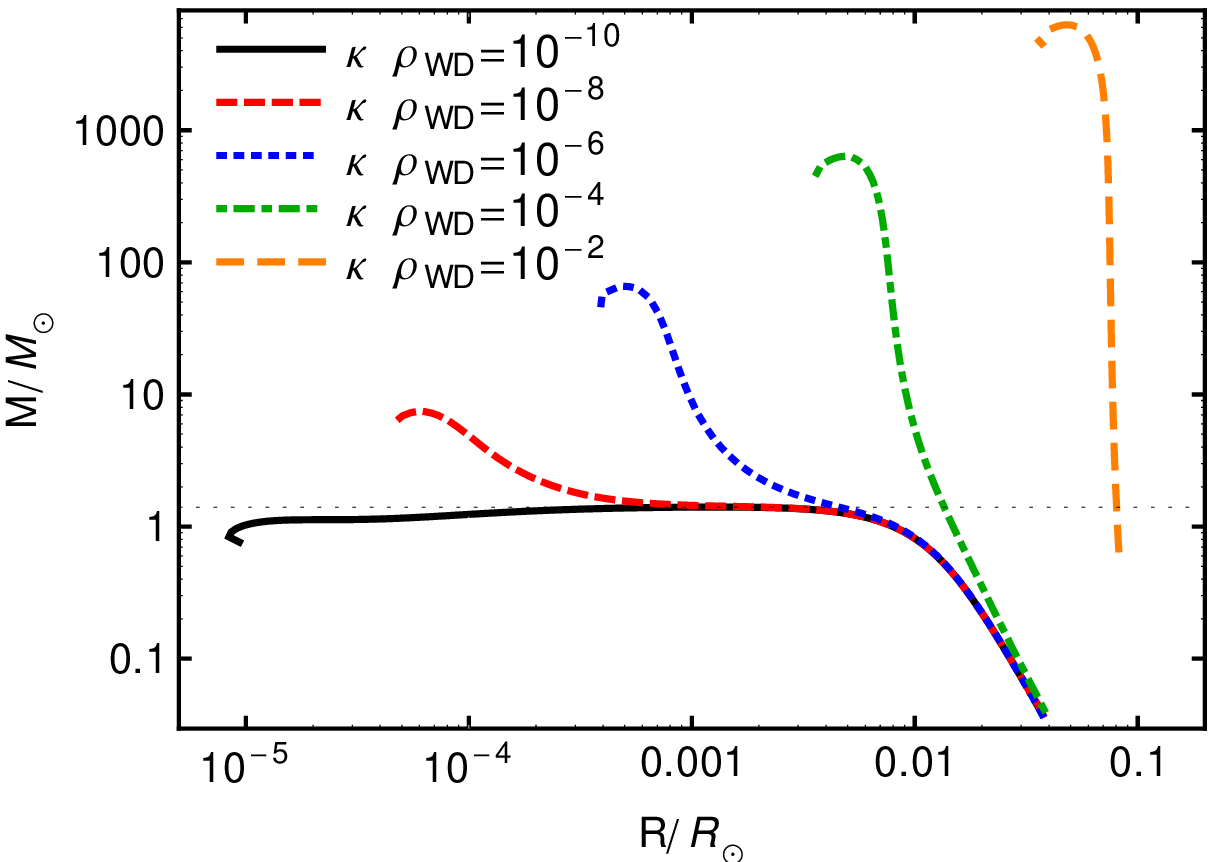,width=6.5cm,angle=0}
\epsfig{file=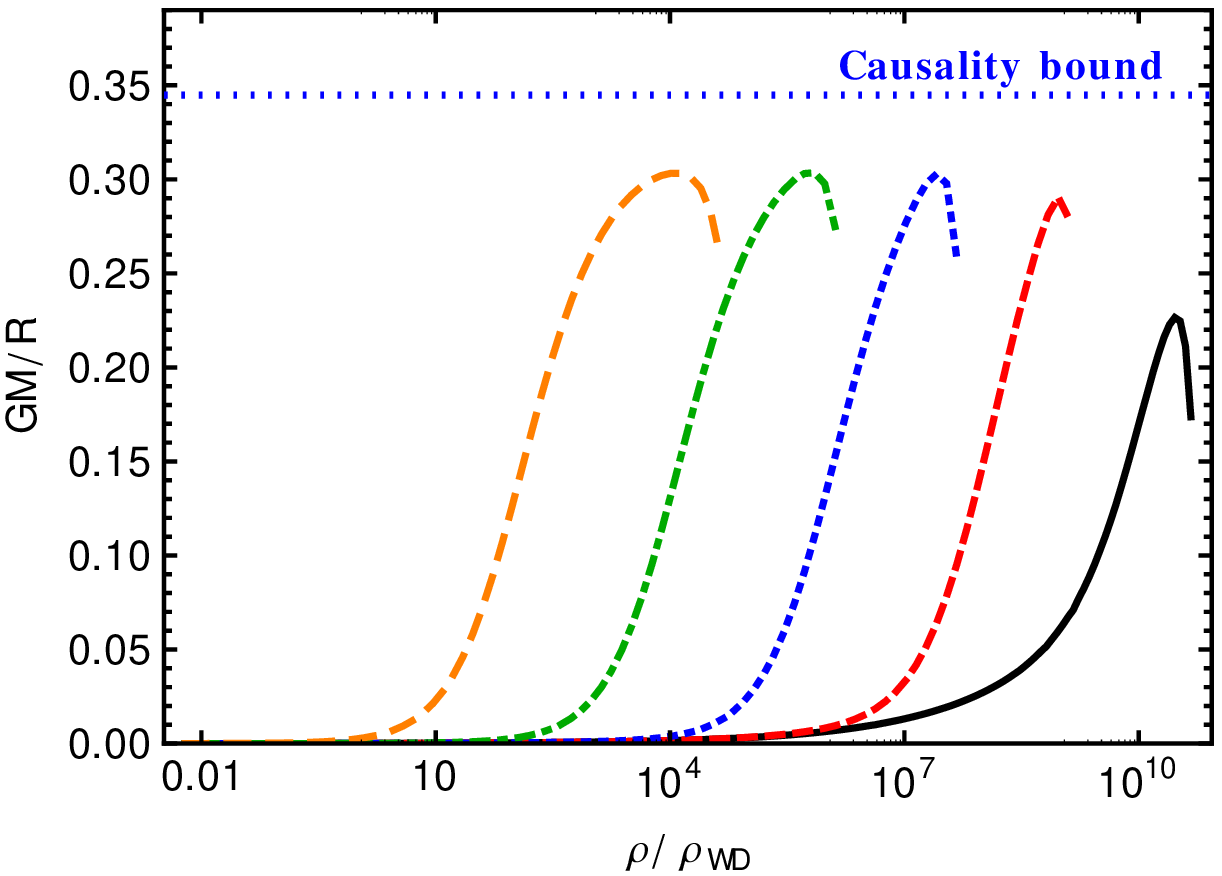,width=6.5cm,angle=0}
\end{tabular}
\caption{Zero-temperature relativistic white dwarfs in EiBI theory. Left panel: mass-radius relation. The horizonal line denotes Chandrasekhar limit, $M\sim1.4M_\odot$. Right panel: compactness as a function of the central density. The horizontal line denotes the maximum compactness allowed by causality, $GM/R\lesssim 0.35$~\cite{Lattimer:2006xb} (cf. also the discussion at the end of Section~\ref{sec:realisticEOS}). Results are expressed in units of a typical white dwarf density, $\rho_{WD}=10^9\text{kg/m}^3$. Curves terminate when condition~\eqref{lim1} is not fulfilled.  
\label{fig:relWD}}
\end{center}
\end{figure*}
%
\subsection{Relativistic zero-temperature white dwarfs}\label{sec:WDrel}
With the modified Tolman-Oppenheimer-Volkoff equation at hand, we can also study relativistic models of zero-temperature white dwarfs, described by the EOS~\eqref{EOS_WD}. Note that this EOS can be explicitly written as $P=P(\rho)$, whereas the inverted function $\rho=\rho(P)$ must be written in parametric form. However, we have already discussed that in EiBI gravity the field equation~\eqref{divT2} also contains $\rho'$ terms, so that it can be used equivalently as an evolution equation for $\rho$ or for $P$. The former choice is more convenient, since we can substitute $P(\rho)$ explicitly in the field equations and solve them for $\rho$ and for the metric fields.

Some results are shown in Fig.~\ref{fig:relWD}.  We focus on $\kappa>0$, since this is the most interesting region of the parameter space. The left panel of Fig.~\ref{fig:relWD} shows the mass-radius relation for a zero-temperature white dwarf in the full EiBI theory. Note that, for any $\kappa>0$, there exists a maximum mass. In the $\kappa\to0$ limit, the classical Chandrasekhar result, $M_{\rm max}\sim 1.4 M_\odot$, is recovered. This is shown in the left panel of Fig.~\ref{fig:relWD} by an dotted horizontal line. 

When $\kappa$ is non-vanishing, the maximum mass can be much larger and it can easily exceed $10^2 M_\odot$. 
Therefore, the absence of white dwarfs with mass $M\gg M_\odot$ can be used to constrain the coupling $\kappa$. Curves in Fig.~\ref{fig:relWD} terminate when the central density is so high that condition~\eqref{lim1} is not fulfilled, but the maximum mass generically exists for smaller central densities. Although the mass can be quite large, the compactness $GM/R$ of the object is limited, as shown in the right panel of Fig.~\ref{fig:relWD}.
\subsection{Realistic nuclear-physics based EOS}\label{sec:realisticEOS}
\begin{figure*}[htb]
 \begin{center}
 \begin{tabular}{cc}
 \epsfig{file=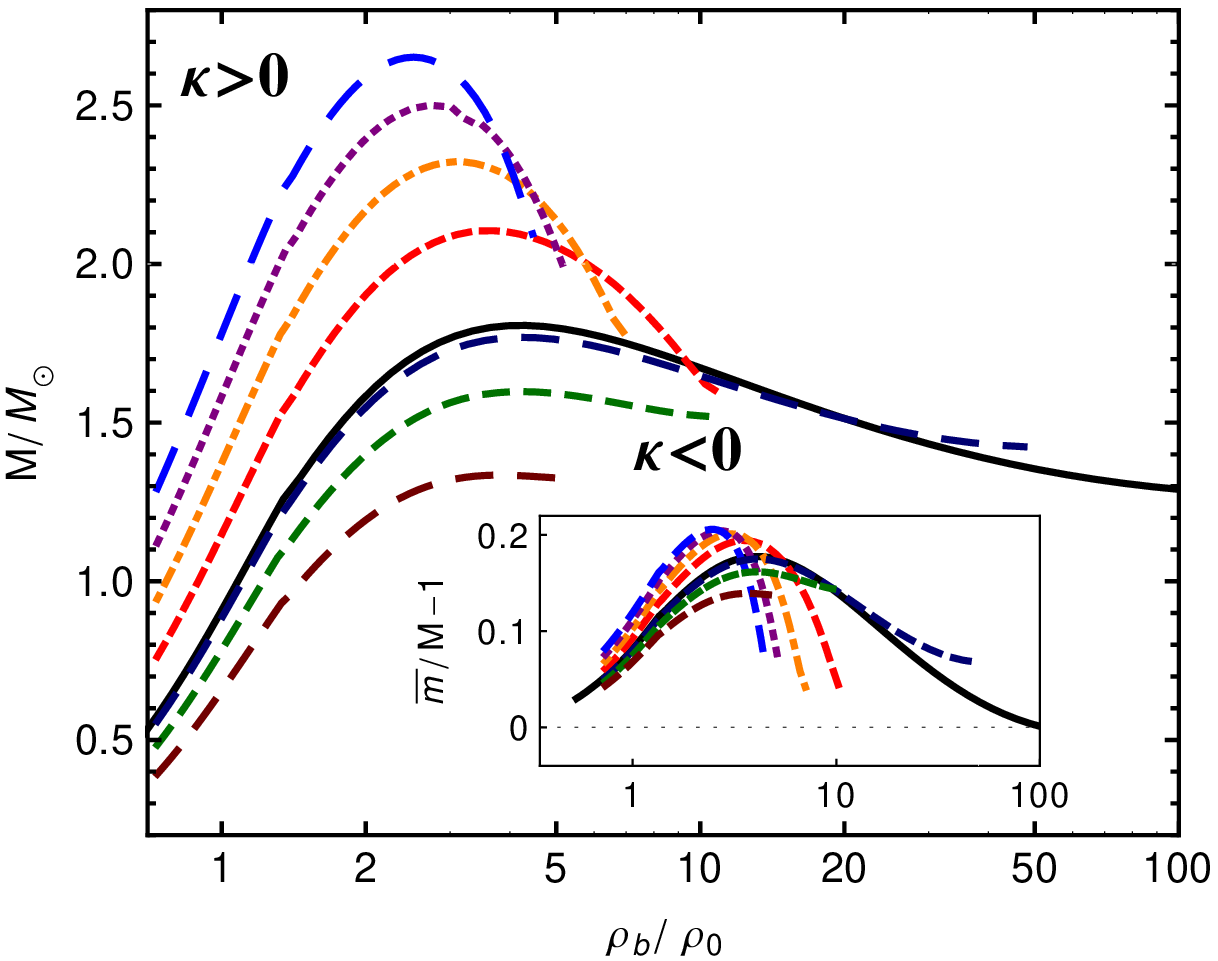,width=6.5cm,angle=0}
 \epsfig{file=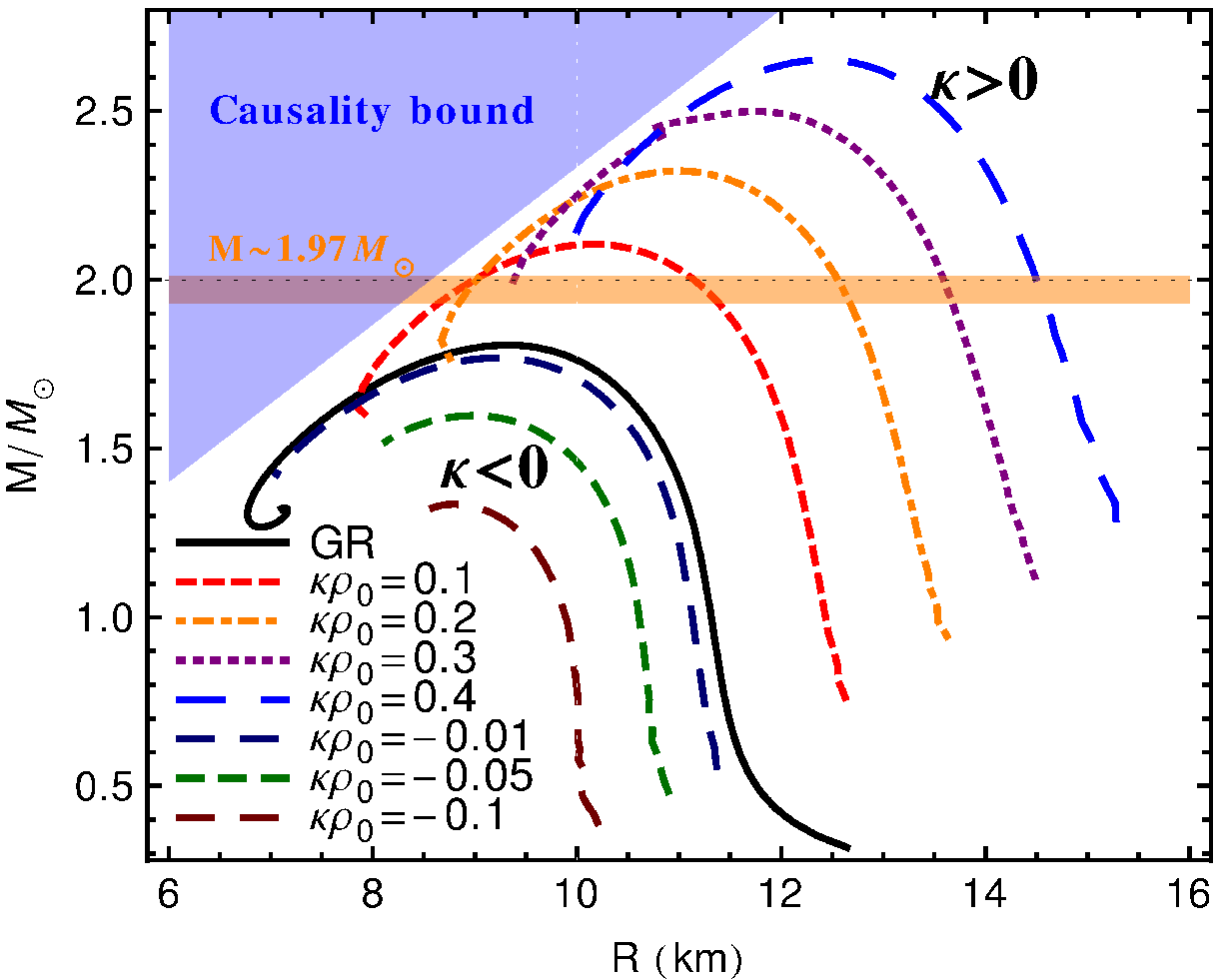,width=6.3cm,angle=0}
 \end{tabular}
 \caption{Compact stars in EiBI theory with FPS EOS obtained by fitting a piecewise polytrope model for different values of $\kappa$. Left panel: mass as a function of the central baryonic density $\rho_b$. Right panel: mass-radius relation. Inset: binding energy as a function of $\rho_b$. Results are normalized by $\rho_0=8\cdot 10^{17}$~kg\ m$^{-3}$, which is a typical central density for NSs. Curves terminate when conditions~\eqref{lim1} or \eqref{lim2} are not fulfilled. The horizontal band denotes the recent observation of a NS with
$M=(1.97\pm0.04)M_\odot$~\cite{Demorest:2010bx}, whereas the shaded region is excluded by causality, $R\gtrsim 2.9 GM$~\cite{Lattimer:2006xb}
 \label{fig:piecewise}}
 \end{center}
 \end{figure*}

Using tabulated, nuclear-based EOS to study NSs in EiBI theory is numerically challenging. The presence of derivatives of the matter fields requires interpolation of the EOS, which can be imprecise, specially close to the radius, where $P\sim0$. 
To avoid this problem, we have used a piecewise polytropic EOS~\cite{Read:2008iy}, which can be constructed analytically as follows. For a set of dividing baryon number densities, $n_0<n_1<n_2...$, we define
\begin{equation}
 \rho(P)=m_b n+\frac{1}{\Gamma_i-1}P\,,\qquad P=K_i(m_b n)^{\Gamma_i}\,,
\end{equation}
for $n_{i-1}<n<n_i$, respectively. Here $m_b n$ is the rest-mass density and $\rho$ is the total energy density. Given a set of $\Gamma_i$ and an initial $P_1=P(m_b n_1)$, the values of $K_i$ are obtained by imposing continuity. In Ref.~\cite{Read:2008iy}, the best fits for $\Gamma_i$ for a three division piecewise are provided for many tabulated EOS. Here, we  shall focus on the FPS EOS, for which~\cite{Read:2008iy}
\begin{equation}
\Gamma_1=2.985,\quad \Gamma_2=2.863,\quad \Gamma_3=2.600\,,
\end{equation}
and $P_1=6.653\cdot10^{34}{\,\text{g}\,}/({\text{cm s$^2$})}$.
Piecewise polytropes can approximated tabulated EOS extremely well.
With the choice above, the fit to the tabulated FPS EOS gives a residual $0.0050$ and an error for the maximum mass and momentum of inertia  of order $0.01\%$. Some results obtained using this EOS are shown in Fig.~\ref{fig:piecewise}.

Notice that, within GR, a standard EOS like FPS would be ruled out by the recent observation of a NS with
$M=(1.97\pm0.04)M_\odot$~\cite{Demorest:2010bx} (denoted by an horizontal band in Fig.~\ref{fig:piecewise}. In EiBI gravity the maximum mass of a NS can be much larger than in GR, thus such observation can be accommodated without invoking a stiffer EOS. 
Curiously, no constraint on $\kappa$ comes from causality, $R\gtrsim 2.9 GM$,~\cite{Lattimer:2006xb}. The latter is always satisfied even for very large values of $\kappa$. This is related to the existence of a maximum compactness, $GM/R\lesssim0.3$, which is independent from the EOS (see e.g. Figs.~\ref{fig:p0MR} and \ref{fig:relWD}). This interesting property have a general interpretation in terms of effective stress-energy tensor~\cite{TerenceJan}.

\subsection{Slowly rotating models}
Slowly rotating stars can be constructed from the corresponding static solutions~\cite{Hartle:1967he}. At first order in the rotation, $g_{t\varphi}=-\zeta(r)r^2\sin^2\theta$, $q_{t\varphi}=-\eta(r)r^2\sin^2\theta$ and the stress-energy tensor for a rotating fluid can be built from from Eq.~\eqref{Tmunu_fluid} with 
\be
u^a=\left\{u^t,0,0,\Omega u^t\right\} \,,\quad u^t=\sqrt{-(g_{tt}+2\Omega_{t\varphi}+\Omega^2 g_{\varphi\varphi})}\,,\nn
\ee
where $\Omega$ is the angular velocity of the fluid. 

Solving the field equation at order ${\cal O}(\zeta,\eta,\Omega)$, we find two equations for $\zeta$ and $\eta$

\begin{eqnarray}
\eta(r)&&=\sqrt{\frac{B p}{F h}} \left[(1 -\kappa  P) \zeta(r)+\kappa  \Omega  A (P+\rho)\right]\,,\nn\\
0&&=4 h^2\left[r^2 \zeta +\left(-r^2+\kappa \right) \eta \right]+r \kappa h' \left(2 \eta +r \eta '\right)\nn\\
&&-\kappa  h \left[r \frac{p'}{p} \left(2 \eta -r \eta '\right)+2  \left(2 \eta +4 r \eta '+r^2 \eta ''\right)\right]\,.\nn
\end{eqnarray}

\begin{figure}[htb]
\begin{center}
\begin{tabular}{c}
\epsfig{file=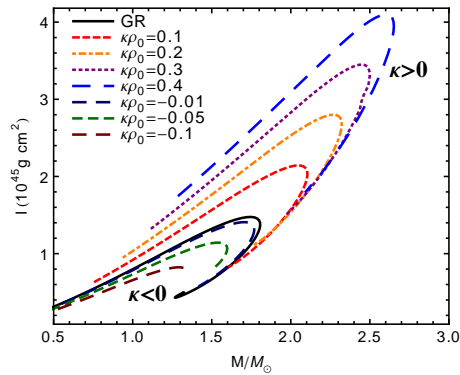,width=6cm,angle=0}
\end{tabular}
\caption{Moment of inertia for FPS EOS obtained with a piecewise polytropic models as a function of the stellar mass for different values of $\kappa$. Results are normalized by $\rho_0=8\cdot 10^{17}$~kg\ m$^{-3}$, which is a typical central density for NSs.
\label{fig:FPS_momInertia}}
\end{center}
\end{figure}

The first equation above is an algebraic relation between $\eta$ and $\zeta$. Substituting this into the second equation above gives a second order ODE for $\zeta$, which has to be solved by imposing regularity of $\zeta$ at the center, $\zeta(0)=\zeta_c$, $\zeta'(0)=0$ and requiring continuity at the stellar radius. At infinity, the asymptotic behavior of $\zeta$ reads $\zeta\to\zeta_\infty+{2J}/{r^3}$, where $J$ is the angular momentum. For asymptotically flat solutions we must impose $\zeta_\infty=0$. In Fig.~\ref{fig:FPS_momInertia} we show the moment of inertia $I=J/\Omega$ as a function of the stellar mass for slowly rotating stellar models obtained with the piecewise FPS EOS discussed above.

Note that Figs.~\ref{fig:piecewise} and \ref{fig:FPS_momInertia} are qualitatively very similar to Figs.~3 and 4 in Ref.~\cite{Pani:2011mg}, which have been obtained using a simple polytropic EOS.
\section{Conclusion}\label{sec:conclusion}
EiBI theory is a viable modified gravity which introduces a highly non-trivial coupling to matter, while being equivalent to Einstein's gravity in vacuum. In this theory, the metric tensor is minimally coupled to matter, thus enforcing the usual stress-energy tensor conservation and geodesic motion. Nevertheless, the non-local Born-Infeld structure of the theory affects the matter sector by introducing higher order corrections in the matter fields and in their gradients.

Even at non-relativistic level, these extra terms allow for a new interesting phenomenology, which is compatible with current experiments. We have shown that in a class of non-relativistic theories --~including the non-relativistic limit of EiBI gravity~-- described by a modified Poisson equation~\eqref{Poisson} with $\kappa>0$, the collapse of non-interacting particles produces a regular, pressureless stars, rather than a singular state. We have discussed the stability of these and other non-relativistic stellar models, showing that pressureless stars are stable against linear radial perturbations and that positive values of $\kappa$ generically tend to stabilize the star. We have also extended original Chandrasekhar's analysis, showing that, for an arbitrarily small value of $\kappa>0$, zero-temperature white dwarfs and NSs do not have a maximum mass, but possess a minimum radius $R\sim \kappa^{1/2}$.

In the fully relativistic EiBI theory, we have constructed several static and slowly rotating compact objects: relativist pressureless stars which do not exist in GR, zero-temperature white dwarfs with relativistic EOS and nuclear-physics base models of NSs. All these solutions have a maximum compactness (always smaller than the causality bound) and a maximum mass, which generically increases as a function of $\kappa$ and can easily be $10-10^3$ larger than in GR. This suggests that, in the fully relativistic theory, the collapse of very massive stars may still form black holes, but this may require collapsing stars of very high mass, depending on the value of $\kappa$.

For these reasons, the most natural and important extension of our work would be to study the relativistic stellar collapse and identify the final state. This would be particularly relevant to understand whether singularities may still be formed during the matter collapse in EiBI gravity, or if their formation is prevented, similarly  to what happens in EiBI cosmology~\cite{Banados:2010ix}.

Near-future experiments (e.g. the proposed NICER mission) will investigate the interior of NSs, allowing for null tests of GR inside matter. Measurements of the NS mass, radius and moment of inertia will also allow to perform tests of alternative theories~\cite{Pani:2011xm} such as EiBI and to constrain the coupling parameters of the theory~\cite{Avelino}. Since NSs are the most extreme matter configurations in the universe, such tests would constrain the matter sector of gravitational theories to unprecedented levels.

\begin{acknowledgments}
  We thank Pedro Gil Ferreira, Jocelyn Read, Jorge Rocha, Ulrich Sperhake and Jan Steinhoff for useful discussions.
  This work was supported by the {\it DyBHo--256667} ERC Starting
  Grant and by FCT - Portugal through PTDC projects FIS/098025/2008,
  FIS/098032/2008, CTE-AST/098034/2008, CERN/FP/123593/2011.  PP acknowledges financial support provided by the European Community through the Intra-European Marie Curie contract aStronGR-2011-298297 and the kind hospitality of the Department of Physics, University of Rome ``Sapienza'', during the last stages of this work.
\end{acknowledgments}
\appendix
\section{Matching conditions}\label{app:matching}
In this appendix, we derive the matching conditions~\eqref{matchingBEI}, focusing on the spherically symmetric case.
In GR, one imposes Darmois-Israel matching conditions~\cite{Israel:1966rt}, i.e.
\begin{equation}
 [g_{ij}]=0\,,\qquad [K_{ij}]=\left[\frac{\partial_r g_{ij}}{\sqrt{g_{rr}}}\right]=0\,,\label{matchingGR}
\end{equation}
across any hypersurface $\Sigma$ of a manifold $({\cal M},g)$. In the formula above, $[...]$ is the jump across the surface, $K_{ij}$ is the extrinsic curvature tensor for $\Sigma$ and $i,j=0,2,3$.
The matching conditions comes from the Einstein equations and the requirement that $\Sigma$ has a well-defined $3-$geometry. For a spherically symmetric spacetimes, starting with a metric ansatz 
\begin{equation}
 ds^2=-F(r)dt^2+B(r)dr^2+r^2d\Omega^2\,,
\end{equation}
the Israel conditions implies $[F]=0=[F'/B]$. 
The first condition can be fulfilled by a time reparametrization, whereas the second one defines the mass of the spacetime. 
Incidentally, these two conditions together with Einstein's equations imply that all the metric functions are $C^1$. Indeed, one can equivalently impose $[F]=[B]=0$.

The same reasoning can be applied to EiBI theory. From the ansatzen~\eqref{ansatzg}-\eqref{ansatzq}, we now have five metric functions $F,B,p,h,A$ and two matter fields, $P$ and $\rho$. Note that, inside matter, $q_{ab}$ is the metric which defines the metric connection $\Gamma$ and, in turn, the covariant derivatives and the curvature tensor $R_{ab}$. Thus, one would expect the mass to be directly related to the auxiliary metric $q_{ab}$, and not to $g_{ab}$.
The junction conditions~\eqref{matchingBEI} follow from the fact that $R_{ab}$ appearing in Eq.~\eqref{eqDIN} is defined in terms of the auxiliary metric $q$. Last of Eqs.~\eqref{matchingBEI} is a consequence of the field equations, while the first two conditions are imposed to ensure that any hypersurface (either of the manifold $({\cal M},g)$ or of $({\cal M},q)$) have a well-defined $3-$geometry. 

The matching conditions~\eqref{matchingBEI} can be derived as follows~\cite{MTW}. Starting from Eq.~\eqref{eqALG}, we perform the integral along the proper distance, $n$, measured perpendicularly through $\Sigma$
\begin{equation}
  \lim_{\epsilon\to0}\int_{-\epsilon}^{\epsilon}dn\sqrt{-q}q^{ab}=\lim_{\epsilon\to0}\int_{-\epsilon}^{\epsilon}dn\lambda\sqrt{-g}\left(g^{ab}-\kappa T^{ab}\right)\,.
\end{equation}
Requiring the absence of discontinuities or delta functions in $g$ (to ensure a well-defined $3-$geometry) and \emph{provided that} $T^{ab}	$ does not contain delta function terms, it follows that $q^{ab}$ is regular at the interface. Hence, from Eq.~\eqref{eqDIN}, we obtain $\lim_{\epsilon\to0}\int_{-\epsilon}^{\epsilon}dn R_{ab}(\Gamma)=0$. Finally, from the decomposition of the four-dimensional Ricci tensor, $R_{ij}^{(4)}$, into the Ricci tensor of the 3-metric, $R_{ij}^{(3)}$,
\begin{equation}
 R_{ij}^{(4)}=R_{ij}^{(3)}-2K_{li}K^l_j+K_l^lK_{ij}-\frac{K_{ij}'}{\sqrt{q_{rr}}}-\frac{\sqrt{q_{rr}}_{,j||i}}{\sqrt{q_{rr}}}\,,
\end{equation}
it follows $[K_{ij}(q)]=0$ to guarantee regularity of $q$.

\bibliography{Eddington}
\end{document}